\documentclass[11pt,a4paper]{article}
\def\runningtitle{Return-decay residuals and tail-risk forecasting}
\usepackage[margin=25mm]{geometry}
\usepackage[T1]{fontenc}
\usepackage{lmodern,microtype,amsmath,amssymb,amsthm,booktabs,tabularx,array,graphicx,xcolor,enumitem,fancyhdr}
\usepackage[hidelinks]{hyperref}
\usepackage{placeins}
\definecolor{ink}{HTML}{17324D}
\definecolor{muted}{HTML}{526475}
\definecolor{wash}{HTML}{EDF3F7}
\newcommand{\E}{\mathbb{E}}
\newcommand{\Prb}{\mathbb{P}}

\newcommand{\ind}{\mathbf{1}}

\newtheorem{proposition}{Proposition}

\theoremstyle{definition}
\setlist{nosep,leftmargin=1.5em}
\newcommand{\sourceurl}[1]{\url{#1}}
\title{Return-Decay Residuals and Tail-Risk Forecasting:\\Timing Artifacts, Conditional Inference, and Cross-Market Evidence}
\author{Chorok Lee\\Korea Advanced Institute of Science and Technology}
\date{}
\hypersetup{pdftitle={Return-Decay Residuals and Tail-Risk Forecasting: Timing Artifacts, Conditional Inference, and Cross-Market Evidence},pdfauthor={Chorok Lee}}
\begin{document}
\maketitle
\begin{abstract}
Can a residual from a fitted return-decay curve predict future factor tail losses, and does it measure crowding? We separate economic edge, overlapping performance statistics, and predictive evidence. A Gaussian counterexample produces a same-month crash-risk ratio of 1.61 without predictable future returns. An eight-factor US reconstruction finds no robust advantage for the tested residual under its original specification. We then freeze and execute a geographic extension with six factors in Europe, Japan, and Asia Pacific excluding Japan, using matched feature baselines, frozen and annually updated forecasts, and separate 2015--2024 and 2025--August 2026 evaluations. Main-era pooled regional log loss increases by 0.00704 with frozen augmentation and 0.00233 with annual refitting. Yet Brier rankings differ, and a matched-history US comparison favors the residual. A controlled diagnostic reproduces the US ranking reversals. A further mechanism study holds observations and labels fixed to isolate the age-dependent candidate family: its effect is secondary in the pooled reversal. Holding A's penalties fixed largely removes the reversal, but reversing the decomposition anchor changes the allocation between history and reselection. The largest changes concentrate in a factor with zero or one validation crash. Complete penalty paths further distinguish changes across feature histories from sensitivity to individual validation months: the zero-event case is stable under all such deletions. Established sequential forecast bounds permit dependent, nonstationary comparisons under an explicitly disclosed research filtration; all 48 intervals include zero and permit nontrivial benefits. An equivalent-basis experiment further isolates regularization effects, with large basis gaps often reflecting poor component forecasts rather than substantial gains over a simple Sharpe baseline. The findings establish sensitivity of the tested evaluation to representation, target definition, and scoring rule, not universal proxy failure, crowding identification, or prospective forecasting value. All forecasts, unfavorable findings, configurations, and provenance are retained.
\end{abstract}

\section{Introduction}
Factor returns can change after a strategy becomes widely known, but return decay does not uniquely identify investor crowding. Changes in risk premia, volatility, implementation costs, sample composition, and chance can generate similar patterns. Evidence on post-publication predictability \cite{mclean} motivates studying competition, while also making it necessary to separate economic mechanisms from descriptive curve fitting.

The earlier manuscript \cite{original} linked a hyperbolic curve to a trading game, fitted the curve to rolling factor Sharpe ratios, and interpreted prediction residuals as crowding. It reported a weak timing strategy and heterogeneous tail-risk associations. This revision retains those research questions while correcting three inferential steps: the equilibrium calculation, the mapping from economic edge to a Sharpe statistic, and the interpretation of contemporaneous residuals as forecasts.

The central question is: \emph{how stable is the incremental forecasting value of a decay residual across markets, target definitions, and matched representations of the same past returns?} A gain against a restricted learner can support a representation choice without identifying crowding. Conversely, an unfavorable score under one reconstruction cannot establish universal failure. The external results and setup diagnostics below make both distinctions consequential. A mechanistic follow-up tests and weakens an initially plausible explanation: resetting the hyperbolic age origin changes the admissible curves, but history and penalty reselection interact in the observed pooled reversal, with the allocation depending on the decomposition anchor.

\paragraph{Contribution and scope.}
This revision separates economic edge, descriptive decay, and predictive evidence; gives a timing counterexample tailored to the proposed residual; and evaluates matched representations in the US and three external geographic universes. It applies established sequential forecast-comparison theory to report what the regional evidence can resolve under dependence and drift. The Cournot-style calculation, conditional-expectation identity, proper scoring rules, and block-resampling principles are established tools. We do not claim a new general learning algorithm or a new general information-theoretic theorem. A controlled computational study separates changed fitting observations, changed candidate families and penalty reselection, retaining the disconfirmation of a dominant origin effect. The reconstruction evaluates one transparent implementation of the question; it is not an exact replication or a test of every possible crowding proxy.

\paragraph{Related evidence.}
Post-publication return declines studied by McLean and Pontiff \cite{mclean} motivate a competition hypothesis without uniquely identifying it. Falck, Rej, and Thesmar \cite{falck} also analyze performance deterioration in relation to overfitting and arbitrage hypotheses. DeMiguel, Mart\'in-Utrera, and Uppal \cite{demiguel} study how trading across characteristics can alleviate price-impact costs; our single-signal model excludes that mechanism. Barroso, Edelen, and Karehnke \cite{barroso} explicitly distinguish feedback assumptions and examine institutional-holdings proxies, finding evidence inconsistent with crowding as an unconditional source of momentum tail risk. A return-curve residual is not the holdings measure in that work. These distinctions preclude treating a similar sign in a residual-based table as a replication of their mechanism.

\section{Measurement and data}
The economic motivation is inverse-linear residual edge $e(t)=a/(N(t)+1)$ under a stylized quantity game, not inverse-linear trader profit. Appendix~\ref{app:game} gives the corrected equilibrium, Poisson-entry calculation, and restrictive bridge to a population Sharpe. It also proves that changing opportunities can reproduce the same edge path without changing participation, and that equal Sharpe ratios need not imply equal tail probabilities. These restrictions motivate treating the fitted residual as a predictive representation, not a participation estimate.

The historical study uses monthly US factor returns from the Kenneth French Data Library \cite{french}: market, size, value, profitability, investment, momentum, and short- and long-term reversal. A replication must archive the exact downloaded series, date range, definitions, units, and data vintage. In particular, the SMB construction must be consistent across merged factor files. The original narrative ends the sample in 2024 while its strategy table is labeled 2017--2025; the input manifest must resolve this discrepancy.

Let $r_{f,t}$ be a monthly factor return: market excess over the risk-free rate for Mkt--RF, and a long--short portfolio-return spread for the other factors. No additional risk-free subtraction is applied to those spreads. A trailing $W$-month Sharpe statistic using conventional square-root-of-time scaling is
\begin{equation}
 \widehat S_{f,t}=\sqrt{12}\,\overline r_{f,t,W}/s_{f,t,W},
\end{equation}
when $s_{f,t,W}>0$. This is $\sqrt{12}$ times the monthly statistic, not generally the Sharpe ratio of annual aggregated returns under serial dependence \cite{lo}. We retain this fixed scaling for comparability and do not interpret it as dependence-correct annual performance. This statistic changes with both mean and volatility. At origin $t$, the tested signal is $C_{f,t}=\widehat m_{f,t|t-1}-\widehat S_{f,t}$, where $\widehat m$ fits the inverse-linear curve to strictly earlier Sharpe observations. A large value is a decay-residual signal. It uses the return at $t$ and can forecast $t+1$, but it cannot be described as forecasting the already observed return at $t$. It is not the equilibrium edge $e(t)$, and an edge model predicts a Sharpe curve only with additional assumptions connecting traded quantities, capital allocation, and risk. Risk-adjusted regression alpha is another estimand and should be separately defined through a benchmark model.

For a descriptive curve comparison, retain signed Sharpe observations and compare the same dates across models. Conditioning on $\widehat S>0$ changes the estimand and can make a positive decaying curve appear more suitable. Because adjacent 36-month Sharpe estimates share 35 monthly returns, ordinary independent-error inference is inappropriate. Resampling should operate on the underlying time series and reconstruct the rolling statistic.

\subsection{A null example for overlapping windows}
Let $r_t$ be independent and identically distributed, with mean zero and variance $\sigma^2\in(0,\infty)$, and let $\bar r_{t,W}=W^{-1}\sum_{j=0}^{W-1}r_{t-j}$. For integer $h\geq0$,
\begin{equation}\label{eq:overlap}
 \operatorname{Corr}(\bar r_{t,W},\bar r_{t+h,W})
 =\frac{\max(W-h,0)}{W},\qquad
 \operatorname{Cov}(\bar r_{t,W},r_{t+1})=0.
\end{equation}
There are exactly $\max(W-h,0)$ shared returns in the covariance sum, each contributing $\sigma^2/W^2$; each mean has variance $\sigma^2/W$. The future one-period return shares no observation and is independent of the trailing window. For $W=36$, successive means have correlation $35/36\approx0.9722$ even though the next return is unpredictable from the past.

This calculation is exact for rolling means, or a statistic standardized by a fixed known volatility. It is not an exact correlation formula for an estimated Sharpe ratio with a random denominator. It demonstrates why persistence or forecastability of overlapping performance statistics does not establish forecastability of nonoverlapping future returns. The attached seeded iid example checks this mechanism only; it is not market evidence.

\paragraph{Vintage and publication availability.}
The French Data Library switched its US research-return production from CRSP FIZ to CIZ beginning with the January 2025 release. The documented monthly-return construction also changed: CIZ compounds daily returns and reinvests dividends on ex-dates rather than month-end \cite{french}. The replication manifest must therefore record the file hash, download date, return convention, and vintage; the switch cannot be inferred merely from a return observation's calendar date.

The research clock and publication clock are distinct. A dated historical factor observation is not proof that the published series was available immediately after that day's or month's close. Unless archived release timestamps or an implementable reconstruction establish availability, a lagged exercise is a chronological research backtest, not a certified real-time trading backtest. Use $\mathcal F_t^{\rm avail}$ for genuinely available inputs and record observation date, release time, and execution time separately.

\section{What a decay residual can and cannot establish}
\subsection{A mechanical same-month crash association}
Consider a deliberately simple null with iid $r_t\sim\mathcal N(0,\sigma^2)$ and a constant zero expected performance benchmark. Define $U_t=-\sqrt W\bar r_{t,W}/\sigma$, with $W\geq2$, and a high-signal state $U_t>k$. Let $z_p=\Phi^{-1}(p)$ define a crash, where $\Phi$ and $\phi$ are the standard normal distribution and density. This construction contains no traders, crowding, or predictable returns.
\begin{proposition}[Timing alone creates a tail association]\label{prop:timing}
For finite $k$ and $p\in(0,1)$, let $\rho=-W^{-1/2}$ and
\[
 J_W(k,p)=\int_{-\infty}^{z_p}\phi(x)
 \left[1-\Phi\left(\frac{k-\rho x}{\sqrt{1-\rho^2}}\right)\right]dx.
\]
The contemporaneous high- and low-signal crash probabilities are
\begin{align}
 \Prb(r_t<\sigma z_p\mid U_t>k)&=\frac{J_W(k,p)}{1-\Phi(k)}>p,\\
 \Prb(r_t<\sigma z_p\mid U_t\leq k)&=\frac{p-J_W(k,p)}{\Phi(k)}<p.
\end{align}
In contrast, both corresponding probabilities for $r_{t+1}$ equal $p$.
\end{proposition}
\begin{proof}
The pair $(r_t/\sigma,U_t)$ is jointly standard normal with correlation $-1/\sqrt W$, since the current return enters the trailing mean. Conditional on $r_t/\sigma=x$, $U_t$ has mean $\rho x$ and variance $1-\rho^2$. Integrating the conditional upper-tail probability gives $J_W$. That upper-tail probability is strictly decreasing in $x$; conditioning $x$ below $z_p$ therefore raises its expectation above its unconditional value. This gives the strict high-state inequality, and total probability gives the low-state inequality. Independence of $r_{t+1}$ from the entire trailing window gives the final assertion.
\end{proof}

At $k=0$, using $-\widehat S_t>0$ instead of $U_t>0$ selects exactly the same dates whenever sample volatility is positive. Thus the conditional-rate formulas at this cutoff also apply to a negative estimated-Sharpe signal under this null, despite its random denominator. This equivalence does not extend to arbitrary nonzero cutoffs or to a fitted nonzero decay curve.

\begin{table}[ht]
\centering\small
\resizebox{\linewidth}{!}{\begin{tabular}{rrrrr}\toprule
$W$ & Exact same-month RR & Simulated same-month RR & Next-month RR & Next-month events\\\midrule
12 & 2.335 & 2.345 & 1.003 & 49809\\
36 & 1.606 & 1.615 & 0.997 & 49806\\
60 & 1.440 & 1.438 & 1.003 & 49806\\
\bottomrule\end{tabular}
}
\caption{Gaussian null with $p=0.10$, high state $-\widehat S_t>0$, and 500,000 simulated monthly observations (seed 20260927). RR is the high/low crash-risk ratio. Same-month exact values follow Proposition~\ref{prop:timing}; next-month population RR is one. Counts, denominators, and software hashes are in \texttt{results/synthetic-timing.json}. These are synthetic observations, not market estimates.}\label{tab:timing}
\end{table}
The diagnostic deliberately uses a known-zero benchmark to isolate timing. It does not reproduce the original fitting procedure or establish how much of the historical association is mechanical. It shows that even a substantial contemporaneous risk ratio is compatible with a no-predictability null. The simulated observations overlap in their signal windows; ordinary iid binomial intervals would not be justified for the state-conditional rates.

\subsection{Information content versus predictive representation}
Let $X_t$ contain every input used to construct the residual, including the trailing returns, the fitted curve parameters or their training history, and the time index. For a fixed deterministic construction, $C_t=g(X_t)$.
\begin{proposition}[No additional information conditional on generating inputs]
For any integrable outcome $Y_{t+1}$,
\[
 \E[Y_{t+1}\mid X_t,C_t]=\E[Y_{t+1}\mid X_t]\quad\text{almost surely}.
\]
Moreover, an unpenalized logistic model already containing $\widehat m_{t|t-1}$ and $\widehat S_t$ has exactly the same set of fitted probability functions after adding $C_t=\widehat m_{t|t-1}-\widehat S_t$ as a linear predictor.
\end{proposition}
\begin{proof}
Measurability of $g$ gives $\sigma(X_t,C_t)=\sigma(X_t)$, proving the conditional-expectation identity. For the logistic model,
\[
 \beta_m\widehat m+\beta_s\widehat S+\theta(\widehat m-\widehat S)
 =(\beta_m+\theta)\widehat m+(\beta_s-\theta)\widehat S.
\]
The augmented and original linear predictors therefore describe the same function class.
\end{proof}
The first statement is a general measurability fact, not a new information-theoretic result. Its value here is to define the paper's empirical interpretation precisely. It does not imply equal finite-sample performance for restricted learners. A nonlinear ratio or a historically fitted trend may be a useful feature for a simple model. Regularization can also change when redundant columns are added, so penalized results require an explicit tuning comparison rather than an appeal to the unpenalized equivalence.

The appropriate hierarchy is: (i) simple trailing-return and volatility controls; (ii) a component-matched model with the curve prediction and estimated Sharpe separately, plus flexible transformations chosen on development data; (iii) the residual representation with comparable tuning; and (iv) independent holdings or flow inputs when available. An improvement at (iii) supports a useful representation under the tested learning procedure. Evidence for a crowding mechanism additionally requires independent measurement or an identification design. Merely adding $C_t$ to a linear model in mean and volatility does not separate an economic mechanism from the nonlinear mean/volatility ratio.

\section{Retrospective public-data reconstruction}\label{sec:reconstruction}
\subsection{Data, estimands, and chronology}
We downloaded the official monthly five-factor, momentum, short-term reversal, and long-term reversal archives on 26 September 2026. All four headers identify the August 2026 CRSP database. The analysis uses July 1963--December 2024, a complete 738-month calendar for all eight factors. Mkt-RF, SMB, HML, RMW, and CMA are taken from one five-factor file to avoid mixing SMB definitions. Values are divided by 100 once; RF is not subtracted again from Mkt-RF or from long-short spreads. SHA-256 hashes, source URLs, and retrieval timestamps are frozen in the repository. Data after December 2024 are not used in this original eight-factor reconstruction; the separately specified extension below uses observations through August 2026.

This is a new-vintage reconstruction. It does not reproduce the original positive-only 1995--2024 fit, original threshold-training window, undocumented preprocessing, or strategy. The evaluation years were already examined in the original research; they are chronologically held out from model fitting here, but are not an untouched scientific confirmation sample. Later-vintage historical revisions also prevent certification of real-time availability.

For each factor, the outcome threshold is its empirical 10th percentile in July 1963--December 1999, using linear quantile interpolation. This threshold stays fixed. Logistic training targets run from January 1975 through December 2009, validation targets from January 2010 through December 2014, and evaluation targets from January 2015 through December 2024. Model weights are refitted on all development targets through 2014 after tuning. Each evaluation prediction uses $C_t$ and other inputs dated no later than $t$ to predict $r_{t+1}$. Thresholds and scalers never use evaluation observations. A separate extra-delay sensitivity predicts the same target months with inputs ending one additional month earlier; it is an availability stress test, not proof of publication timing.

The primary descriptive estimand is the equal-factor mean evaluation log-loss difference
\[
 \widehat\Delta=\frac{1}{8T}\sum_{f=1}^8\sum_{u=1}^T
 \left\{\ell(Y_{f,u},\widehat p^{\rm aug}_{f,u})
             -\ell(Y_{f,u},\widehat p^{\rm ctrl}_{f,u})\right\},\quad T=120,
\]
where $\ell(y,p)=-y\log p-(1-y)\log(1-p)$ and negative differences favor augmentation. This weights factor-month forecasts equally, not investment capital. Log loss and Brier loss are established proper scores \cite{gneiting}; their use here does not introduce a new scoring criterion.

\subsection{Curve estimation and prediction models}
The primary window is 36 months. At each origin $t$, fit the hyperbola to every finite signed Sharpe statistic ending strictly before $t$, requiring at least 60 such observations. Set age zero at the first complete window and measure age in years. Search $\lambda\in\{0,0.01,0.03,0.1,0.3,1,3\}$ per year; for each value, estimate nonnegative amplitude by least squares. The minimizing past-sample squared error determines $\widehat m_{t|t-1}$. Zero decay is allowed, so a fitted constant is a valid outcome, not silently discarded. A sensitivity allows signed amplitude. Constant, exponential, linear, and one-step persistence forecasts of the rolling statistic are compared on the same dates.

Separate ridge logistic models are fit for each factor. The control model contains the latest available return, trailing mean, trailing standard deviation, and calendar age. The augmented model adds $C_t$. The component model adds $\widehat m_{t|t-1}$ and $\widehat S_t$ separately. Flexible controls include the Sharpe ratio, squares of return/mean/volatility/Sharpe, and a mean--volatility interaction in addition to the controls. A historical-event-rate predictor uses $(n_1+1/2)/(n+1)$ on development labels. No class reweighting or outcome-based test-period selection is used.

For each nonconstant model, standardize features using training values and minimize mean log loss plus $\alpha\|\beta\|_2^2/2$, with an unpenalized intercept. Select $\alpha\in\{0.001,0.01,0.1,1,10\}$ by validation log loss separately for each factor/model, then refit through 2014 with newly fitted development scalers. Components and residuals have different regularization geometries and dimensions even though they share generating information. Their comparison evaluates the specified learning procedures, not unrestricted conditional information content. All tuning losses and fitted coefficients are retained.

\subsection{Dependence and multiplicity}
We resample aligned monthly vectors of paired loss differences using circular blocks, preserving both temporal runs and contemporaneous cross-factor dependence. The primary block length is 12 months, with 6, 24, and 36 as sensitivity values, and 4,000 resamples with seed 20260926. This follows the established block-bootstrap rationale for dependent observations \cite{kunsch}; it is not an iid bootstrap of factor-month rows. The macro resampling range is percentile-based. For the eight primary factor contrasts, joint ranges use the 95th percentile of the maximum absolute centered bootstrap mean divided by each factor's bootstrap standard error. This family does not include all sensitivity/model comparisons, which remain exploratory.

The realized $\widehat\Delta$ is exactly known for the retained forecasts and outcomes. We therefore report the bands as \emph{descriptive resampling ranges}, not established 95\% confidence intervals for that finite sample or a future era. Under suitable stationarity and weak-dependence conditions, a conditional population target could be $\Delta=\E[\ell(Y,p^{\rm aug})-\ell(Y,p^{\rm ctrl})\mid\text{development information}]$. Those conditions are not established for this recursively constructed, drifting series. Models, thresholds, and hyperparameters are not refitted inside this loss bootstrap. The nominal resampling quantiles consequently do not establish coverage, learning-procedure uncertainty, or absence of a practically useful effect.

\subsection{Measured results}
\begin{table}[ht]
\centering\small\begin{tabular}{lrr}\toprule
Model & Log loss & Brier score\\\midrule
Historical event rate & 0.47148 & 0.14435\\
Return/volatility controls & 0.47104 & 0.14438\\
Controls + residual & 0.47340 & 0.14468\\
Controls + curve + Sharpe & 0.48397 & 0.14569\\
Flexible return controls & 0.47571 & 0.14561\\
\bottomrule\end{tabular}

\caption{New-vintage reconstruction, January 2015--December 2024. Equal-factor mean scores over 960 factor-month forecasts; lower is better. These are new measured results, not the original study's numbers.}\label{tab:models}
\end{table}
Adding the residual worsens average log loss by 0.00236. The primary 12-month-block resampling range is approximately $[0.00038,0.00459]$. This small deterioration is specific to the frozen-model protocol. A favorable comparison with a weaker fitted model does not rescue a general value claim: the historical-event-rate and basic-control predictors have lower average losses than the augmented model.

\begin{figure}[ht]
\centering\includegraphics[width=\linewidth]{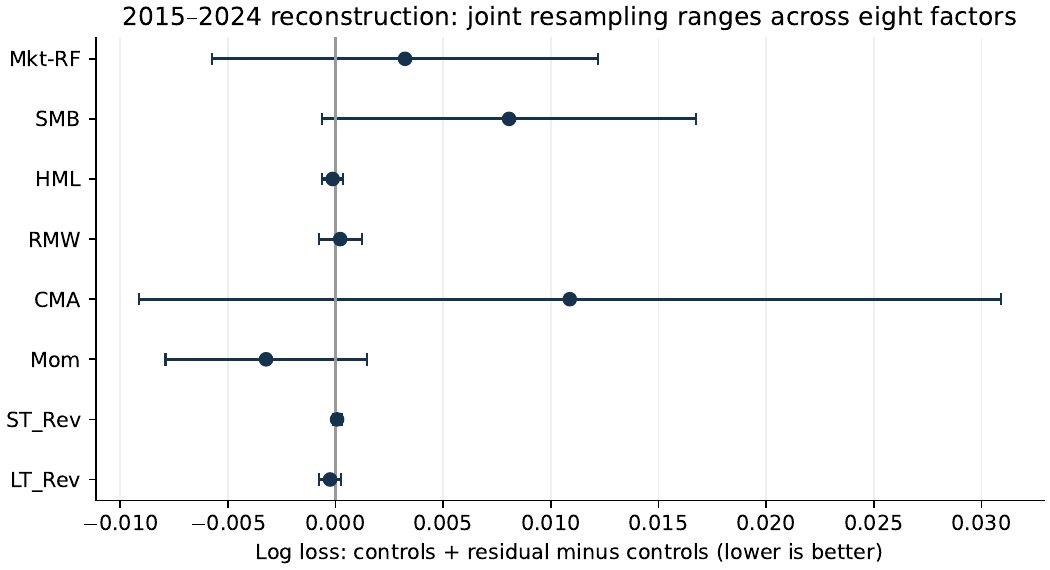}
\caption{Primary factor contrasts and joint resampling ranges across eight factors, using joint 12-month calendar blocks. Every range contains zero; isolated pointwise gains are not treated as confirmed factor-specific findings.}\label{fig:forecast}
\end{figure}

\begin{table}[ht]
\centering\small\begin{tabular}{lrrrr}\toprule
Factor & Events/120 & Predicted rate & Observed rate & $\Delta$ log loss\\\midrule
Mkt-RF & 14 & 0.119 & 0.117 & +0.00323\\
SMB & 15 & 0.105 & 0.125 & +0.00806\\
HML & 22 & 0.102 & 0.183 & -0.00013\\
RMW & 20 & 0.111 & 0.167 & +0.00022\\
CMA & 19 & 0.072 & 0.158 & +0.01088\\
Mom & 21 & 0.140 & 0.175 & -0.00323\\
ST\_Rev & 21 & 0.138 & 0.175 & +0.00007\\
LT\_Rev & 31 & 0.121 & 0.258 & -0.00025\\
\bottomrule\end{tabular}

\caption{All eight factors: event counts, augmented-model mean forecast probability, realized frequency, and augmented-minus-control log loss. Each factor has 120 evaluation months.}\label{tab:factors}
\end{table}
Event frequencies exceed the historical 10\% threshold probability for every factor and reach 31/120 for long-term reversal. Mean probabilities substantially understate several realized frequencies; these averages assess calibration-in-the-large only, not full conditional calibration. There is no basis here for claiming calibrated operational risk forecasts.

The fitted hyperbola loses to persistence on rolling-Sharpe mean squared error for all eight factors (Appendix~\ref{app:additional}). This comparison concerns an overlapping performance statistic, not future return predictability. Market and profitability fits choose zero decay at every primary evaluation origin, demonstrating why an enforced narrative of decay across factors would misdescribe even the fitted model.

\subsection{Sensitivity and failure-case interpretation}
\begin{table}[ht]
\centering\small\begin{tabular}{lrr}\toprule
Specification & $\Delta$ log loss & 95\% block range\\\midrule
Primary: 36 months & +0.00236 & [+0.00038, +0.00459]\\
annual refit & +0.00115 & [-0.00012, +0.00246]\\
extra delay1 & +0.00002 & [-0.00091, +0.00076]\\
signed amplitude & +0.00236 & [+0.00038, +0.00459]\\
window12 & -0.00033 & [-0.00504, +0.00428]\\
window60 & +0.00118 & [-0.00084, +0.00302]\\
\bottomrule\end{tabular}

\caption{Macro augmented-minus-control log-loss differences. All descriptive resampling ranges use 12-month blocks. Window, extra-delay, and signed-amplitude runs were listed in the reconstruction configuration; annual refitting was added after the primary result as an explicitly exploratory response to calibration drift.}\label{tab:sensitivity}
\end{table}
No sensitivity establishes a robust gain. For annual expanding refitting, the validation-selected penalties remain fixed while model weights and scalers are refit each January using targets observed through the preceding December. This reduces the primary deterioration to roughly 0.00115, with a resampling range containing zero. The prediction ledger records each actual fit cutoff. Signed amplitude produces the same primary predictions, indicating that the nonnegative-amplitude constraint is not responsible for this particular result. These outcomes limit this feature construction; they do not prove that independent holdings-based crowding measurements lack predictive value.

\paragraph{Calendar concentration and forecast drift.}
A further post-results diagnostic holds every primary forecast fixed and reports all ten calendar years and both five-year halves (Appendix~\ref{app:temporal}). Omitting any single year's evaluation rows leaves a positive mean loss difference, ranging from 0.00179 to 0.00270. Thus no single omitted year reverses the observed average ranking. However, about 64\% of the summed deterioration occurs in 2023--2024, and augmentation improves the annual score in 2016, 2020, and 2021. The result is heterogeneous rather than a uniform penalty in every period.

Across 2015--2019 and 2020--2024, the observed event frequency rises from 66/480 (13.75\%) to 97/480 (20.21\%), whereas the augmented model's mean probability changes from approximately 10.96\% to 11.79\%. These marginal discrepancies reinforce the limits on operational calibration and stationary-bootstrap interpretation. The calendar analysis is descriptive: removing evaluation rows does not refit models or remove those observations' influence on later rolling features. It supplies neither independent confirmation nor a new significance test.

\subsection{Matched representation ablations}
To isolate the hyperbolic choice from the Sharpe transformation, a post-results ablation adds exactly one feature to the same controls: raw Sharpe, a constant-fit residual, or a linear-fit residual. The latter two use the same strictly past signed observations as the hyperbola. Each model has the same validation-selected ridge grid, chronology, and scaling procedure. Recomputed controls and hyperbolic predictions exactly match all 960 original reconstruction rows.
\begin{table}[!ht]
\centering\small\begin{tabular}{lrr}\toprule
Model & Log loss & $\Delta$ vs. hyperbolic\\\midrule
Controls & 0.47104 & -0.00236\\
Controls + hyperbolic residual & 0.47340 & +0.00000\\
Controls + Sharpe & 0.47066 & -0.00274\\
Controls + constant residual & 0.47176 & -0.00164\\
Controls + linear residual & 0.47216 & -0.00124\\
\bottomrule\end{tabular}

\caption{Exploratory matched ablation on the same 2015--2024 outcomes. Every augmented model adds one feature. Lower loss is better; no result is independent confirmation.}\label{tab:matched}
\end{table}
Controls plus Sharpe have loss 0.47066, compared with 0.47340 for the hyperbolic residual. The paired Sharpe-minus-hyperbolic difference is $-0.00274$, with a descriptive 12-month block range $[-0.00489,-0.00063]$. Its difference from basic controls is only $-0.00039$, with a range spanning zero. Thus the comparison does not establish a robust useful signal, but it provides no observed advantage for the hyperbolic specification. Constant and linear residuals also have lower observed loss than the hyperbolic residual.

\subsection{Conditional-design learnability and adaptive nulls}
The failed empirical comparison cannot establish that the learner would detect every useful signal. We hold the momentum feature history fixed and generate synthetic labels with probability $\operatorname{logit}^{-1}(\alpha+\beta z(C_t)+\delta\ind\{t+1\geq201501\})$. Here $z$ uses pre-2015 scaling, $\alpha$ gives mean development probability 0.10, $\beta\in\{0,0.5,1\}$, and $\delta\in\{0,0.8\}$. All six conditions were recorded before label generation, with 100 replicates each. Each replicate repeats train/validation selection and refitting for the basic and augmented models. These labels are artificial Bernoulli responses on a fixed design, not observed or simulated factor crashes; this is not end-to-end power or coverage validation.
\begin{table}[!ht]
\centering\small\resizebox{\linewidth}{!}{
\begin{tabular}{rrrrrr}\toprule
$\beta$ & Test shift & Mean $\Delta$ & Wins & MC 95\% interval & Oracle $\Delta$\\\midrule
0 & 0 & +0.00078 & 52/100 & [0.423, 0.615] & -0.00326\\
0 & 0.8 & +0.00014 & 49/100 & [0.394, 0.587] & -0.04874\\
0.5 & 0 & -0.00014 & 59/100 & [0.492, 0.681] & -0.00521\\
0.5 & 0.8 & +0.00192 & 46/100 & [0.366, 0.557] & -0.05021\\
1 & 0 & -0.00414 & 80/100 & [0.711, 0.867] & -0.01316\\
1 & 0.8 & +0.01154 & 39/100 & [0.300, 0.488] & -0.03359\\
\bottomrule\end{tabular}
}
\caption{Conditional-design synthetic positive control. $\Delta$ is augmented-minus-control test log loss; a win means $\Delta<0$. Wilson intervals quantify Monte Carlo uncertainty in the win frequency across independent label draws, not market-data uncertainty. Oracle $\Delta$ compares generating probabilities with controls.}\label{tab:positive}
\end{table}
The strong stable signal ($\beta=1,\delta=0$) yields 80/100 wins and mean difference $-0.00414$. With the same signal and a test intercept shift of 0.8, the mean difference becomes $+0.01154$, with only 39/100 wins. This provides a concrete failure case: model comparison can deteriorate despite an imposed signal when calibration shifts. Moderate effects remain difficult to distinguish here. We therefore retain the limited empirical conclusion rather than asserting signal absence or equivalence within a chosen practical margin.

\paragraph{Sequential return positive control.}
To go beyond a fixed feature design, a second experiment generates all 738 monthly returns recursively:
\[
 r_t=0.1-\beta\tanh(C_{t-1})+\sigma_t\epsilon_t,
 \qquad \epsilon_t\overset{\rm iid}{\sim}\mathcal N(0,1).
\]
Here $C_{t-1}$ is the actual chronologically fitted residual, set to zero while unavailable; $\beta\in\{0,0.2,0.5\}$, with $\sigma_t=1$ before 2015 and either 1 or 1.5 thereafter. These artificial units and feedback dependence are chosen to plant predictability, not to model structural crowding. Fifty independent paths per condition repeat return generation, pre-2000 threshold estimation, rolling statistics, decay fitting, ridge selection, and final forecasting. All 300 paths finish; online and batch residual calculations agree within $2.0\times10^{-15}$.
\begin{table}[!ht]
\centering\small\resizebox{\linewidth}{!}{
\begin{tabular}{rrrrrr}\toprule
$\beta$ & Test $\sigma$ & Mean $\Delta$ & Wins & MC 95\% interval & Oracle $\Delta$\\\midrule
0 & 1 & +0.00034 & 21/50 & [0.294, 0.558] & -0.00352\\
0 & 1.5 & -0.00032 & 30/50 & [0.462, 0.724] & -0.06804\\
0.2 & 1 & -0.00015 & 27/50 & [0.404, 0.670] & -0.00608\\
0.2 & 1.5 & +0.00185 & 27/50 & [0.404, 0.670] & -0.11840\\
0.5 & 1 & -0.00240 & 31/50 & [0.482, 0.741] & -0.01853\\
0.5 & 1.5 & -0.00485 & 34/50 & [0.542, 0.792] & -0.11449\\
\bottomrule\end{tabular}
}
\caption{Sequential-return simulation, with all six conditions retained. $\Delta$ is augmented-minus-control log loss; Wilson intervals describe Monte Carlo uncertainty in win frequencies. The oracle knows the generating conditional mean, test volatility, and frozen event threshold; its advantage includes privileged knowledge unrelated to the residual. Wins are not calibrated test power.}\label{tab:sequential}
\end{table}
For $\beta=0.5$ without a scale shift, the mean difference is $-0.00240$ (Monte Carlo standard error 0.00087), with 31/50 wins. With the scale shift, it is $-0.00485$ (standard error 0.00607), with 34/50 wins. The latter mean is imprecise, and the signs do not establish that shifting volatility improves the comparison. Null and moderate-signal conditions show small, mixed differences. The oracle contrast is negative even at $\beta=0$, because controls must estimate probabilities and do not know the generating scale shift. This experiment closes the joint-pipeline implementation check under the stated generators, while leaving general detection power, resampling coverage, and empirical usefulness unestablished.

We also pass 400 complete synthetic return paths through the actual recursive hyperbolic fitting and state-selection pipeline: 100 each from iid Gaussian, standardized $t_5$, centered standardized lognormal, and stochastic-volatility generators. Each has 738 months and no crowding mechanism. The volatility process has predictable conditional variance, so it is not a null of unpredictable tail risk. Crash thresholds use observations through 1999; state cutoffs are the median or 75th percentile of pre-2015 fitted residuals. All comparisons score the same 120 target months, retaining every path and state count.

Across these settings, mean same-month high-minus-low crash-rate differences range from 3.5 to 6.8 percentage points. Next-month sample differences have broad between-path ranges spanning zero. Empty states occur in 1--10 of 100 paths per setting; their undefined differences are retained as missing and counted, and summaries condition on defined differences. These distributions characterize finite-sample behavior of adaptive states and random denominators, not tests of the original result or confidence intervals. The full table is in Appendix~\ref{app:nulls}.

\section{Geographic transport and conditional forecast uncertainty}\label{sec:external}
\subsection{Frozen external evaluation}
After completing the US investigation, we froze an external protocol and archived six additional official French Data Library files before computing regional forecast outcomes. The three universes are Europe, Japan, and Asia Pacific excluding Japan; each contributes Mkt--RF, SMB, HML, RMW, CMA, and regional WML, explicitly aliased to Mom. The provider-defined country lists are disjoint, but the series share global and currency shocks, dates, and a data provider. They are not statistically independent replications. Their 202608 archive headers identify Bloomberg data, unlike the US archives' CRSP provenance. Returns are simple USD returns; regional market returns already subtract the US one-month Treasury-bill return \cite{frenchregional}.

The common input calendar is December 1990--August 2026, with 429 complete months. Five-factor files begin in July 1990; momentum documentation begins in December, while archived momentum rows begin in November. We explicitly exclude that earlier row from analysis while preserving the raw archive and apply the documented common December start to every factor. Each factor's event threshold is its 10th percentile over the 109 observations through December 1999. Training targets are 2000--2009 (120 months), validation targets 2010--2014 (60), and evaluation eras are 2015--2024 (120) and January 2025--August 2026 (20). The short second era is separately reported and is historical, not prospective confirmation.

The 36-month window, 60-observation curve warmup, decay grid, ridge grid, six forecast models, and all transformations remain fixed. The matched models are prevalence, controls, and controls plus one of the hyperbolic residual, raw Sharpe, constant residual, or linear residual. Frozen fits use development targets through 2014. Annual fits retain the same validation-selected penalties, but update coefficients, scalers, and prevalence every January using only targets through the previous December. An additional six-factor US replay uses the same common history and splits, giving a descriptive geography comparator without confounding region with a longer fitting history. All region--factor rows, forecasts, cutoffs, coefficients, and adverse outcomes are retained.

\subsection{An established bound with a disclosed research filtration}
The earlier block ranges rely on conditions that drift can compromise. For the external study we instead adapt established sequential forecast-comparison theory \cite{choeramdas,howard}. This is not a new confidence-sequence method. Its target differs from the stationary population quantity sometimes associated with a bootstrap interval.

Let $\mathcal R$ contain the completed US data analysis, protocol-selection information, and external metadata inspected before freezing the regional study. For each era, $\mathcal F_0$ contains $\mathcal R$ and the external panel history before that era. Thereafter $\mathcal F_t$ adds the complete external return panel through month $t$, including its derived features and crash labels. In particular, some US observations in $\mathcal R$ are later than a historical regional forecast month and can be informative about it. Conditioning on this information is mathematically legitimate but changes the target: it does not recover a real-time historical information set or turn geographic evidence into independent confirmation. The second era resets the sequence with the first era's external history already included.

For two $\mathcal F_{t-1}$-measurable probabilities $p_{0,f,t},p_{1,f,t}$, clipped to $[10^{-12},1-10^{-12}]$, write their paired log-loss contrast as
\[
 d_{f,t}(y)=y\log\frac{p_{0,f,t}}{p_{1,f,t}}
 +(1-y)\log\frac{1-p_{0,f,t}}{1-p_{1,f,t}},\qquad
 D_t=F^{-1}\sum_{f=1}^F d_{f,t}(Y_{f,t}).
\]
The conditional estimand is
\[
 \Theta_n=\frac1n\sum_{t=1}^n\E[D_t\mid\mathcal F_{t-1}],
\]
with positive values favoring forecaster 0. It is a running average of conditional expected scores along this sequence, not a constant future-era performance parameter. The realized score average is already known exactly. Predictable endpoints are
\[
 a_t=F^{-1}\sum_f\min\{d_{f,t}(0),d_{f,t}(1)\},\qquad
 b_t=F^{-1}\sum_f\max\{d_{f,t}(0),d_{f,t}(1)\}.
\]
Thus the monthly width is $c_t=b_t-a_t=F^{-1}\sum_f|\operatorname{logit}(p_{0,f,t})-\operatorname{logit}(p_{1,f,t})|$. The square of this full-panel width is used; no independence gain across factors or regions is assumed.

\begin{proposition}[Finite-grid application of sequential forecast bounds]\label{prop:cs}
For $K$ fixed comparisons and $J$ fixed positive values $\lambda_j$, let $Q_n=\sum_{t\leq n}c_t^2$ and
\[
 h_n=\min_{1\leq j\leq J}\left\{\frac{\log(2KJ/\alpha)}{\lambda_j}+\frac{\lambda_jQ_n}{8}\right\}.
\]
Then, simultaneously across these comparisons and all $n\geq1$, the intervals
\[
 I_n=\left[\overline D_n-h_n/n,\overline D_n+h_n/n\right]
 \cap[\overline a_n,\overline b_n]
\]
cover their respective $\Theta_n$ with probability at least $1-\alpha$, under the stated predictability and boundedness conditions. Stationarity and serial or cross-factor independence are not required.
\end{proposition}
\begin{proof}
Set $\mu_t=\E[D_t\mid\mathcal F_{t-1}]$. Conditional Hoeffding's lemma gives
$\E[\exp\{\lambda(D_t-\mu_t)\}\mid\mathcal F_{t-1}]\leq\exp(\lambda^2c_t^2/8)$.
Hence $\exp\{\lambda\sum_{t\leq n}(D_t-\mu_t)-\lambda^2Q_n/8\}$ is a nonnegative supermartingale starting at one. Apply Ville's inequality with error $\alpha/(2KJ)$ to each sign, grid value, and comparison, and take a union bound. Every already-adjusted boundary holds together, so minimizing over the grid is valid. Each $\mu_t\in[a_t,b_t]$ justifies the support intersection. This is the standard concentration argument specialized to the paired binary score and whole-month panel.
\end{proof}

We fixed $\alpha=0.05$, $\lambda_j=2^j$ for $j=-6,\ldots,10$, and $K=48$: four strata (three regions and their equally weighted pool), three contrasts (hyperbolic--controls, Sharpe--controls, hyperbolic--Sharpe), two fitting modes, and two eras. The pool gives equal weight to each of 18 series. Family overlap requires no independence assumption. No unadjusted factorwise confidence claims, extra start dates, or all-era combined interval are added. A finite grid supplies the claimed time-uniform bound but is not asserted to be asymptotically sharp. It pays for selection within this grid, not undeclared researcher search. Historical release-time availability remains unverified.

\subsection{Measured scores, uncertainty, and counterexamples}
The complete run produces 5,040 external prediction rows (18 series, 140 months, two fitting modes) and 1,680 separate US comparison rows, with no failed series. Table~\ref{tab:external-models} reports every matched model for the main era. Hyperbolic augmentation changes pooled regional log loss from 0.30419 to 0.31123 with frozen fits and from 0.29941 to 0.30174 with annual fits. Raw Sharpe and a constant residual have lower observed main-era log loss than the hyperbolic residual in both modes. These are fixed-protocol comparisons; annual refitting was included before external evaluation.
\begin{table}[!ht]
\centering\small\begin{tabular}{lrrrr}\toprule
 & \multicolumn{2}{c}{Frozen} & \multicolumn{2}{c}{Annual}\\
Model & Log loss & Brier & Log loss & Brier\\\midrule
Prevalence & 0.30201 & 0.08560 & 0.30003 & 0.08527\\
Controls & 0.30419 & 0.08673 & 0.29941 & 0.08543\\
Hyperbolic residual & 0.31123 & 0.08608 & 0.30174 & 0.08545\\
Sharpe & 0.30228 & 0.08572 & 0.29879 & 0.08507\\
Constant residual & 0.30137 & 0.08561 & 0.29826 & 0.08517\\
Linear residual & 0.30565 & 0.08585 & 0.29933 & 0.08522\\
\bottomrule\end{tabular}

\caption{Pooled external scores, 2015--2024, across 18 series. All nonprevalence alternatives add their named single feature to the same four controls. Lower scores are better. Log loss is primary; Brier scores are descriptive secondary outcomes.}\label{tab:external-models}
\end{table}
The ranking is score-specific. Frozen hyperbolic augmentation slightly improves Brier loss, from 0.08673 to 0.08608, despite increasing log loss; its extension-era Brier difference is also negative, $-0.00100$. Two proper scoring rules need not order imperfect forecasts alike. We therefore make no claim of forecast dominance across scores.

Table~\ref{tab:external-resolution} shows the conditional intervals and the improvement magnitude still compatible with each lower endpoint. All 48 final intervals, and every one of their 3,360 saved prefixes, include zero. For the main pooled hyperbolic--controls comparison, the frozen bound is $[-0.03003,0.06172]$ and the annual bound is $[-0.01830,0.03229]$. These ranges permit nontrivial beneficial effects as well as deterioration. They neither establish practical equivalence nor prove lack of useful signal. ``Gain not excluded'' is $\max(0,-L)$ for lower endpoint $L$; it measures inferential resolution in natural-log-loss units, not an economic materiality threshold or predicted investment return.
\begin{table}[!ht]
\centering\small\resizebox{\linewidth}{!}{\begin{tabular}{lllrrrr}\toprule
Era & Fit & Pair & $\widehat\Delta$ & Lower & Upper & Gain not excluded\\\midrule
2015--24 & Frozen & H$-$C & +0.00704 & -0.03003 & +0.06172 & 0.03003\\
2015--24 & Frozen & S$-$C & -0.00191 & -0.01063 & +0.00681 & 0.01063\\
2015--24 & Frozen & H$-$S & +0.00895 & -0.01986 & +0.05633 & 0.01986\\
2025--26 & Frozen & H$-$C & +0.02436 & -0.02468 & +0.25549 & 0.02468\\
2025--26 & Frozen & S$-$C & +0.00630 & -0.00947 & +0.03241 & 0.00947\\
2025--26 & Frozen & H$-$S & +0.01806 & -0.02093 & +0.23296 & 0.02093\\
2015--24 & Annual & H$-$C & +0.00233 & -0.01830 & +0.03229 & 0.01830\\
2015--24 & Annual & S$-$C & -0.00062 & -0.00782 & +0.00658 & 0.00782\\
2015--24 & Annual & H$-$S & +0.00294 & -0.01535 & +0.02857 & 0.01535\\
2025--26 & Annual & H$-$C & +0.00050 & -0.01372 & +0.06649 & 0.01372\\
2025--26 & Annual & S$-$C & +0.00314 & -0.01725 & +0.01923 & 0.01725\\
2025--26 & Annual & H$-$S & -0.00263 & -0.01208 & +0.06196 & 0.01208\\
\bottomrule\end{tabular}
}
\caption{Pooled external score contrasts and 95\% simultaneous conditional confidence-sequence endpoints for the frozen 48-comparison family. H, S, C denote hyperbolic augmentation, Sharpe augmentation, and controls. The target conditions on prior US research information and sequential external returns, not historical real-time information. The extension ends in August 2026.}\label{tab:external-resolution}
\end{table}
Geographic and temporal exceptions matter. Hyperbolic--controls log-loss differences are positive in all three regions during 2015--2024, but Japan's annual-fit extension difference is $-0.00182$. The pooled annual extension also favors hyperbolic over Sharpe augmentation by $-0.00263$, reversing their main-era order. These comparisons retain wide intervals and are not promoted to discovered subgroups. All regional and secondary-score results appear in Appendix~\ref{app:external}.

The matched-history US benchmark supplies a stronger counterexample to a blanket negative conclusion. Its frozen main-era hyperbolic--controls difference is $-0.04739$, and its annual difference is $-0.00608$. The frozen extension difference is $-0.13071$, whereas annual refitting gives $+0.00920$. This benchmark changes feature history, threshold reference, training span, and the factor set relative to the original eight-factor reconstruction. Its ranking reversal cannot be assigned to geography alone or treated as independent confirmation. It motivates the explicitly post-results setup diagnostic below.

\paragraph{A controlled, explicitly post-results setup diagnostic.}
To investigate the US reversal, we retain the same six factors and 2015--2024 targets while crossing two choices: curve-feature history starting in July 1963 or December 1990, and threshold reference starting at those dates and ending in December 1999. All four cells train on 2000--2009 and validate on 2010--2014 with the same penalty grid and selection rule. A fifth reference restores 1975--2009 training with full histories. This diagnostic was specified after seeing the reversal and is descriptive; no extra confidence claim is added to the external family. Its 3,600 prediction rows exactly recover both earlier US implementations on their common six factors.
\begin{table}[!ht]
\centering\small\resizebox{\linewidth}{!}{
\begin{tabular}{lrrrrr}\toprule
Setup & Events & Controls & Hyperbolic & Sharpe & H$-$C\\\midrule
Full features / full threshold & 111/720 & 0.49010 & 0.48450 & 0.48421 & -0.00561\\
Short features / full threshold & 111/720 & 0.49010 & 0.53010 & 0.48421 & +0.03999\\
Full features / short threshold & 112/720 & 0.52183 & 0.50799 & 0.50716 & -0.01384\\
Short features / short threshold & 112/720 & 0.52183 & 0.47443 & 0.50716 & -0.04739\\
Legacy training reference & 111/720 & 0.44310 & 0.44627 & 0.44264 & +0.00317\\
\bottomrule\end{tabular}
}
\caption{Post-results US setup diagnostic. Full history starts July 1963; short history starts December 1990. The four crossed cells share 2000--2009 training; the reference starts in 1975. Lower log loss is better. Threshold changes redefine labels, so comparisons across threshold rows concern different event targets.}\label{tab:history}
\end{table}
With the full threshold reference, shortening curve-feature history changes the hyperbolic--controls difference from $-0.00561$ to $+0.03999$; under the short threshold reference it changes from $-0.01384$ to $-0.04739$. Controls and raw-Sharpe predictions remain unchanged within each threshold comparison, to numerical precision. Their trailing windows agree and the translated age covariate is absorbed by training standardization; the residual procedure additionally depends on the curve's fitting history and age origin. The interaction includes validation-selected penalty responses, not a structural effect of older observations alone. Shortening classifier training from 1975 to 2000 with full feature and threshold histories improves the relative contrast, but worsens both absolute losses (controls 0.44310 to 0.49010; hyperbolic 0.44627 to 0.48450). A more favorable relative score can therefore reflect a deteriorating comparator.

Threshold references change 23 of the 720 labels, although total event counts move only from 111 to 112. CMA supplies a substantial counterexample: its hyperbolic--controls difference is $+0.29031$ for short features with the full threshold (19 events) and $-0.23051$ with the short threshold (11 events). This attribution is descriptive and does not select a new winning factor or protocol. The diagnostic demonstrates why a single pooled ranking cannot separate representation sensitivity from target-definition sensitivity.

\paragraph{Implementation validation under drift and dependence.}
A separate synthetic check generated 2,000 independent paths of 240 months with six tasks: iid or lag-Markov probabilities with an abrupt shift, crossed with independent or perfectly shared within-month innovations. Forecasts used only past outcomes and deterministic signals, so the running conditional score contrast was known. For each 500-path condition, both the $K=1$ and $K=48$ budgets gave zero failures and zero anytime misses (pathwise-coverage Wilson Monte Carlo interval $[0.9924,1]$). The $K=48$ check evaluates one pair with that allocation, not a simulated 48-pair family's coverage. Prefix perturbations, exact replay, and duplicate-task invariance also pass. These conservative diagnostics check implementation; the proof supplies coverage under its assumptions, and simulation does not establish historical admissibility.

\section{Separating fitting history from the admissible curve family}\label{sec:mechanism}
The setup comparison above changes both the observations used to fit a curve and the origin of its age coordinate. We now separate those changes. This diagnostic was selected after the US reversal; its complete specification was pushed before computing the new cell forecasts. It reuses the same historical outcomes, so it is an explanatory experiment on the learning procedure rather than independent confirmation or an additional hypothesis test.

\subsection{Coordinate transport and an implicit decay constraint}
For a curve $f_{A,\lambda}(x)=A/(1+\lambda x)$, with $A,\lambda\geq0$ and positive denominators at every fitted and forecast age, let $x'=x-c$ and $c>0$. Elementary reparameterization gives
\begin{equation}\label{eq:transport}
 \lambda'=\frac{\lambda}{1+c\lambda},\qquad
 A'=\frac{A}{1+c\lambda},\qquad
 f_{A,\lambda}(x)=f_{A',\lambda'}(x').
\end{equation}
For the same fitting observations, each candidate basis vector is multiplied by $1+c\lambda$. Its least-squares amplitude, including projection onto nonnegative amplitudes, is divided by that factor. Every candidate's fitted values, SSE and forecast are therefore unchanged. Transporting the ordered rate grid preserves the selected curve, with identical tie-breaking; when every amplitude is zero, the selected rate has no unique substantive interpretation. The residual is unchanged, and translation of the age control is removed by training standardization. The subsequent penalized logistic objective is consequently identical when targets and tuning rules agree.

Keeping the same numerical rate grid after translating the origin generally changes the candidate curves. This is not only a coarse-grid issue: $\lambda\in[0,\infty)$ maps to $\lambda'\in[0,1/c)$. Here the origins are separated by $c=329/12$ years and $1/c=0.03647$ per year; the native short-history rates 0.1, 0.3, 1 and 3 are outside the transported continuous domain. The two finite grids need not be nested. For one fixed, nonzero curve at age $a>0$,
\begin{equation}\label{eq:decay-cap}
 -\frac{d}{da}\log f_{A,\lambda}(a)=\frac{\lambda}{1+a\lambda}<\frac1a,
 \qquad
 \frac{f_{A,\lambda}(a+h)}{f_{A,\lambda}(a)}\geq\frac{a}{a+h},\quad h\geq0.
\end{equation}
An old origin thus caps admissible local decline even with a continuous nonnegative-rate search. This bound concerns extrapolation along a \emph{single fixed curve}; it does not bound changes in the sequentially refitted $m_t$.

Time-origin transformations of linear, exponential and rational curves are established \cite{gluzman}; the distinction between a nonlinear model's predictions and its parameter coordinates is also established \cite{transtrum}. We use the algebra as an explanatory and software-consistency check, not a new general theorem. Such checks are related to established metamorphic regression testing \cite{luu}. Transport preserves a declared reference family; it does not show that the earlier origin is economically correct or improves prediction. A dataset's starting date is not an observed strategy birth date.

\subsection{A fixed-target mechanism diagnostic}
All cells use the six common US factors, 2000--2009 training, 2010--2014 validation and 2015--2024 evaluation. Each of the two threshold references is retained as a separate task. Cell A uses full curve-fitting history and its original age coordinate; B uses the short history on that same coordinate; C uses the identical short observations with the native short-history origin and unchanged rate grid; D uses C's coordinate with B's transported candidate family. Thus A to B changes fitting observations, B to C changes the represented family, and B to D checks exact equivalence. Exponential, unrestricted-intercept linear and constant residuals serve as invariance controls: at fixed observations their fitted curves do not change with the origin. The exponential amplitude absorbs the translation without changing its rate grid.

The first analysis selects the ridge penalty separately in each cell. The second holds the numerical penalty at the value selected in A for each factor, target definition and representation, while still refitting coefficients and scalers. It isolates penalty reselection within the procedure, not all downstream adaptation. Differences along A to B to C telescope at fixed labels, but are an ordered procedure decomposition, not a unique causal attribution. No extra inferential family or economic effect is claimed.

\paragraph{The origin hypothesis is secondary in the measured comparison.}
The full diagnostic contains 11,520 forecast rows and 69,120 model probabilities, reusing 720 distinct factor--month return outcomes under two label definitions. Every A/C reference forecast exactly recovers the earlier diagnostic. B/D hyperbolic curve forecasts agree within $1.4\times10^{-15}$ and classifier probabilities within $4.2\times10^{-12}$. Exponential, linear, constant, controls and raw-Sharpe B/C probabilities agree within $5.5\times10^{-15}$. Independent raw-archive and least-squares reconstruction, using a separate Newton classifier solver, reproduces all selected penalties; the largest probability discrepancy from the original L-BFGS-B solution is $1.4\times10^{-6}$, with objective gap at most $1.2\times10^{-11}$. These numerical checks establish implementation agreement, not statistical confirmation.

\begin{table}[!ht]
\centering\small
\begin{tabular}{llrrrrr}\toprule
Threshold & Penalty & A H$-$C & B H$-$C & C H$-$C & B$-$A & C$-$B\\\midrule
Full & Selected & -0.00561 & +0.03840 & +0.03999 & +0.04401 & +0.00159\\
Full & A penalty & -0.00561 & -0.00328 & -0.00340 & +0.00233 & -0.00012\\
Short & Selected & -0.01384 & -0.04743 & -0.04739 & -0.03360 & +0.00004\\
Short & A penalty & -0.01384 & -0.01001 & -0.01148 & +0.00383 & -0.00147\\
\bottomrule\end{tabular}

\caption{Fixed-target US mechanism diagnostic, mean log loss over six factors and 120 months. H$-$C is hyperbolic augmentation minus controls; negative values favor augmentation. ``A penalty'' retains each factor's A-selected numerical ridge while refitting coefficients/scalers. Last two columns give ordered changes; they are not unique causal components. D recovers B to numerical precision.}\label{tab:mechanism}
\end{table}
Under selected penalties, shortening fitting observations at the unchanged origin raises log loss by 0.04401 under the full threshold and lowers it by 0.03360 under the short threshold. The subsequent native-origin change adds only 0.00159 and 0.00004, respectively. Holding A's penalties instead yields small, positive observation-history changes under both thresholds, 0.00233 and 0.00383. The candidate-family change is real, but it does not explain the principal pooled reversal. Transporting the family preserves the chosen benchmark and does not repair the dominant measured effect or establish a better forecaster.

Factor heterogeneity remains important. In the full-threshold selected analysis, CMA's A-to-B increase is 0.26684, exceeding the pooled increase after division by six because other factors partly offset it. Its selected ridge changes from 0.01 to 0.001. Under the short threshold, CMA's ridge instead changes from 0.001 to 0.01 and its loss falls by 0.21268. CMA's 60-month validation segment contains only one event under the full threshold and zero under the short threshold; development contains nine and seven events. This is observed validation support, not proof that proper-score tuning is invalid. The fixed-penalty comparison shows that penalty reselection is consequential for these complete learning procedures. It does not establish a universal rare-event failure law or recommend A's penalty for future use.

A small pooled origin contrast can also conceal material opposing effects: full-threshold C minus B changes CMA loss by $+0.02277$ and SMB loss by $-0.01117$. SMB changes its selected penalty from 1 to 0.001; its largest individual forecast-probability change is 0.133. Thus ``secondary in the pooled reversal'' does not mean every factor's predictions are invariant. The interpretation remains score-specific: selected short-threshold A-to-B log loss improves by 0.03360 while Brier loss worsens slightly, by 0.00000866. Under the full threshold, B/C hyperbolic augmentation has worse log loss but better Brier loss than controls. All family, factor, score and event/non-event decompositions are retained, including the unfavorable and favorable cases. These experiments explain a specific historical procedure comparison; further data and methodological evidence would be needed for a general ML contribution.

\subsection{Complete ridge paths and a second falsification}
The first mechanism result motivates a second, separately frozen diagnostic. We evaluate all five declared ridge penalties for every factor, representation, target definition and A/B/C cell. Training fits produce the 60 validation forecasts; development refits produce the 120 evaluation forecasts. For each validation month, we remove only its contribution to candidate mean loss and recompute the selected penalty, without refitting the training models. These 60 overlapping perturbations describe selection sensitivity; they are not independent observations, confidence estimates or a new deployed rule. The study retains 194,400 model probabilities and 12,960 deletion choices. All 51,840 matched parent probabilities and original validation losses are recovered exactly.

\begin{table}[!ht]
\centering\small
\begin{tabular}{llrrrr}\toprule
Threshold & Cell & Events & Ridge & Validation gap & Deletion switches\\\midrule
Full & A & 1/60 & 0.01 & 0.002037 & 1/60\\
Full & B & 1/60 & 0.001 & 0.001560 & 1/60\\
Full & C & 1/60 & 0.001 & 0.000495 & 1/60\\
Short & A & 0/60 & 0.001 & 0.003817 & 0/60\\
Short & B & 0/60 & 0.01 & 0.003960 & 0/60\\
Short & C & 0/60 & 0.01 & 0.002840 & 0/60\\
\bottomrule\end{tabular}

\caption{CMA hyperbolic-model validation support and selection sensitivity. The gap is runner-up minus minimum mean validation log loss. Each deletion removes one validation month's loss from all candidate averages; training fits remain fixed. CMA is shown because the complete earlier diagnostic found concentration there. Other factors, representations and every fixed-penalty path are retained.}\label{tab:ridge-support}
\end{table}
This follow-up rejects a stronger explanation equating sparse events with local selection instability. Under the short threshold, all 60 CMA validation outcomes are zero, yet every single-month deletion preserves the selected penalty in each cell. A selects 0.001 and B/C select 0.01: the winner changes across feature histories while remaining stable under these within-cell deletions. Under the full threshold, deleting the sole CMA event reverses the selected penalty in each cell; all 59 non-event deletions preserve it. Full-threshold SMB is more deletion-sensitive, changing its selection on 4/60, 47/60 and 55/60 deletions in A, B and C. Event support, sensitivity to individual validation months and sensitivity to feature construction are distinct diagnostics.

\begin{figure}[!ht]
\centering\includegraphics[width=\linewidth]{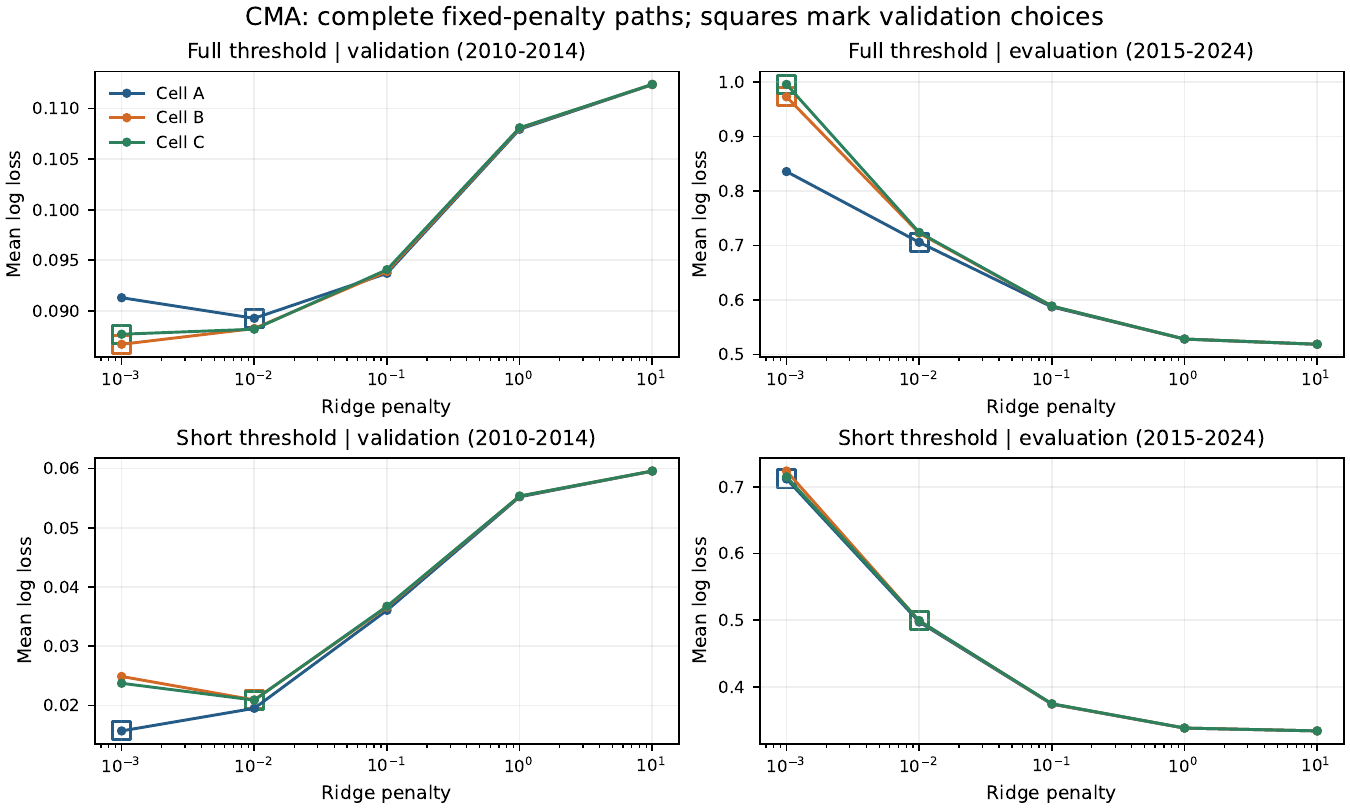}
\caption{Complete CMA hyperbolic-model ridge paths, displayed after concentration was observed. Squares mark the original validation choices in both columns; evaluation losses do not select a replacement rule. Validation fits use 2000--2009 targets, whereas evaluation fits use 2000--2014. Therefore the columns differ in fitting sample as well as evaluation period and event support. The two threshold rows concern different outcomes.}\label{fig:ridge-path}
\end{figure}
The large history response is not unique to the hyperbola. Under short-threshold selected penalties, exponential and constant augmentation change log loss by $-0.03367$ and $-0.03410$ from A to B, close to the hyperbolic $-0.03360$. Exact origin invariance therefore does not protect against the observed fitting-history and tuning interaction. Nor does local deletion stability certify transportability to the later evaluation period. These are concrete limitations of this tested analysis. The data do not validate a general repair, justify an evaluation-selected penalty, or establish a newly discovered universal instability mechanism.

\subsection{Selection margins and order-dependent attribution}
A third post-results diagnostic uses all existing candidate paths, without selecting a replacement predictor. Let $a_j$ and $b_j$ denote mean validation losses for penalty $j$ before and after a history change, and let $i$ minimize $a_j$ with the declared first-candidate tie rule. Put $m=\min_{j\ne i}(a_j-a_i)$ and $d_j=b_j-a_j$. The elementary sufficient condition
\[
m>\max_j d_j-\min_j d_j
\]
preserves the selected penalty. It is invariant to adding a common shift to all losses. Failure of this bound does not imply a switch. The exact retrospective condition checks every $b_j-b_i$, requiring strict positivity for earlier-priority candidates and nonnegativity for later ones; it reconstructs a known selection rather than predicting an unseen failure.

Across all six representations, six factors, two target definitions and three directed transitions (A to B, B to C, A to C), 190 of 216 selections remain unchanged. The sufficient condition certifies 113; 77 additional selections are stable despite its failure. All 26 switches fail the sufficient condition, as the algebra requires. Of the 113 certificates, 72 come from invariant controls or Sharpe augmentation; hyperbolic augmentation is certified in only 3 of 36 comparisons, although 30 are stable. These are overlapping procedure comparisons, not independent diagnostic-validation examples or a statistical accuracy estimate. For CMA hyperbolic augmentation, A-to-B changes the selected penalty under both targets, whereas B-to-C stability is certified under both. Neither the bound nor local deletion stability certifies future forecast quality.

To expose attribution dependence, let $T_h(j)$ be evaluation loss for history $h$ and numerical penalty $j$, with its coefficients and scaler refitted on the declared development sample. If $i$ and $k$ are the two validation choices, the total change has two exact decompositions:
\begin{align*}
T_b(k)-T_a(i)&=[T_b(i)-T_a(i)]+[T_b(k)-T_b(i)]\\
&=[T_a(k)-T_a(i)]+[T_b(k)-T_a(k)].
\end{align*}
The first changes representation before selection; the second reverses that order. For full-target hyperbolic A-to-B mean log loss, the total is $+0.044011$. The first allocation is $+0.002327$ to the fixed-penalty history contrast and $+0.041684$ to reselection. The reverse allocation is $+0.021620$ to reselection and $+0.022391$ to the history contrast. Thus a claim that tuning uniquely explains nearly all the reversal would be incorrect. All 432 score decompositions, including Brier loss, are retained. These identities sharpen a historical case study; they do not establish a new stability theorem, a causal mediation effect, or an independently validated repair.

\subsection{Equivalent feature bases and penalty geometry}
A fourth diagnostic isolates the role of the penalty using two bases of the same linear predictor class. This is a different estimand from the earlier scalar-residual augmentation: both new models contain the same controls plus two columns, either $(S,m)$ or $(S,C)$ with $C=m-S$. Four curve constructions (hyperbolic, exponential, linear, constant), all six factors, A/B/C histories, both fixed targets and all five penalties were specified before computing this diagnostic. The same training, validation and evaluation calendars are retained; these are repeatedly studied historical outcomes, not independent confirmation.

Write the raw basis relation as $X_{SC}=X_{Sm}B$. With separately centered/scaled designs including an intercept, $Z_{SC}=Z_{Sm}T$, where $T$ is recomputed on each fitting sample. If $P=\operatorname{diag}(0,I)$ is the component-model penalty, the equivalent residual-coordinate penalty is $T^\top PT$: coefficients obey $\beta=T\gamma$ and hence $\beta^\top P\beta=\gamma^\top T^\top PT\gamma$. The intercept remains unpenalized. Constant columns are retained with scale one; dropping a column constant in training could alter later predictions. Ridge transformation sensitivity and sparse-sample tuning problems have established precedents \cite{sinkovec,cawley}; this linear reparameterization algebra is not a new theorem. The experiment independently optimizes both objectives rather than merely mapping stored fitted coefficients.

All 5,040 fits succeed. The transported models recover the component probabilities with maximum absolute difference $2.97\times10^{-13}$. A separate implementation verifies all 648,000 retained probability entries, including repeated baseline entries, against 453,600 unique recomputed probabilities; the maximum discrepancy is $1.43\times10^{-10}$. Native isotropic ridge selects different penalties in 62 of 144 factor--construction--history--target comparisons. The native residual basis improves mean evaluation log loss in 77 comparisons and worsens it in 67. No basis is universally preferable.

A frozen, explicitly subjective feasibility rule required consequential concordant log-loss/Brier changes in at least three factors in a setup, plus pooled magnitude and breadth across constructions and targets. Thirteen of 24 setups qualify and all four constructions meet the breadth rule. This warrants an unused-task study under that resource-allocation rule; it is not a significance result or evidence of methodological originality.

The strong-comparator result limits the practical interpretation. For A under the full target, the four native residual models have mean log loss between 0.48393 and 0.48402, close to controls plus Sharpe at 0.48421. Component-model losses range from 0.54272 to 0.59972. Thus a large basis gap can mainly reflect avoiding a poor comparator, with little additional benefit over the simple baseline. Penalty transport restores the reference estimator, including its unfavorable scores; it is not a demonstrated performance repair. Table~\ref{tab:basis} retains every setup, including opposite-direction and scoring-rule counterexamples. The result supports a consequential representation/regularization interaction in this case study, while leaving transfer and useful new curve information unestablished.

\FloatBarrier
\section{A rebuilt external ETF activity study}
The missing original ETF files do not prevent a new study. We rebuilt a measurement pipeline using daily Yahoo Finance price and share-volume observations \cite{yahoo}, frozen with retrieval timestamps and hashes. Issuer objectives motivate two thematic mappings, momentum--MTUM and value--VLUE, with IVV as a broad-market activity control \cite{etfs}. Both factor ETFs began in April 2013; partial April activity is excluded. The remaining six French factors receive no invented ETF mappings. Long-only thematic funds are not identical to the long--short French factors, and these two pairs do not validate the original factor taxonomy.

For ETF $j$, define $D_{j,t}=n_t^{-1}\sum_{d\in t}P_{j,d}V_{j,d}$ and $g_{j,t}=\log D_{j,t}-\log D_{j,t-1}$. Closing price times share volume is an approximation to traded dollars, not actual transaction dollars or fund flows. We use the close field rather than dividend-adjusted close; the archived responses report no split events. All three funds share 2,938 selected daily dates and 140 full months from May 2013 to December 2024. Missing or unequal calendars are rejected rather than filled. Comparing with IVV detects relative omissions but cannot detect a vendor omission shared by all three funds.

The frozen primary analysis uses 138 targets from July 2013 to December 2024. We regress $g_{j,t+1}$ on $C_{f,t}$ and controls: lagged own and IVV activity growth, ETF log-price return, factor return and volatility, market excess return and its absolute value, and linear time. Coefficients are reported per full-overlap standard deviation of the residual; all slopes are full-sample partial associations, not held-out forecasts. Joint circular row blocks refit every coefficient, with 4,000 draws and block lengths 6, 12, 24, and 36. Scaling is held fixed; the two primary slopes define the primary family. Because stationarity and coverage are unverified, bands are descriptive resampling ranges.
\begin{table}[!ht]
\centering\small\begin{tabular}{llrrr}\toprule
Specification & Pair & Slope & Lower & Upper\\\midrule
primary & Mom--MTUM & -0.0115 & -0.0839 & +0.0608\\
primary & HML--VLUE & -0.0110 & -0.2102 & +0.1883\\
same month & Mom--MTUM & -0.0366 & -0.1065 & +0.0332\\
same month & HML--VLUE & -0.0284 & -0.2010 & +0.1443\\
crossmapped & Mom--VLUE & +0.0084 & -0.1830 & +0.1998\\
crossmapped & HML--MTUM & +0.0181 & -0.0571 & +0.0933\\
share volume & Mom--MTUM & -0.0108 & -0.0817 & +0.0601\\
share volume & HML--VLUE & -0.0132 & -0.2096 & +0.1833\\
difference signal & Mom--MTUM & -0.0329 & -0.1727 & +0.1068\\
difference signal & HML--VLUE & -0.1068 & -0.3378 & +0.1243\\
\bottomrule\end{tabular}

\caption{New ETF observations: residual slopes in log activity-growth units per signal standard deviation. Lower/Upper are joint two-slope descriptive ranges at block length 12. Secondary families are exploratory. Both mapped primary ranges and all displayed secondary ranges span zero.}\label{tab:etf}
\end{table}
Primary slopes are $-0.01153$ for Mom--MTUM and $-0.01095$ for HML--VLUE, with wide ranges. Share-volume outcomes, differenced signals, and crossmapped pairs do not establish a stable association. Crossmappings are specificity comparisons, not guaranteed negative controls. The same-month variant moves the signal, IVV activity, and return/volatility controls to contemporaneous dates while keeping own activity lagged; it does not isolate a signal-timing change alone. All leave-year-out slopes and block-length results are retained.

Exchange trading does not measure net creations/redemptions or investor ownership: the issuer distinguishes secondary trading from authorized-participant transactions \cite{etfprospectus}. Therefore neither a positive nor a null result in this rebuilt activity study identifies crowding or its absence. It closes the missing-observation gap for the newly specified activity question, while replacing the undocumented historical correlation with a transparent, separately labeled test.

\section{Limitations and conclusion}
The corrected game supplies a restrictive economic motivation, while the Gaussian null demonstrates a mechanical same-month association. Neither identifies crowding. A residual computed from past returns concerns representation conditional on its generating history. Shared shocks, overlapping windows, limited independent temporal information and revised data vintages constrain inference and historical replication.

Forecast rankings vary across markets, target definitions and scores. The controlled US follow-up shows that age-dependent candidate families have a small pooled effect, while fitting history and penalty reselection interact, with large changes concentrated in CMA. The allocation between history and reselection depends on the decomposition anchor; CMA has zero or one validation event. A second diagnostic separates history sensitivity from deletion sensitivity: the zero-event case is stable under every single-month deletion. Exact family transport preserves a chosen model class; no general repair or forecasting improvement is established. These findings explain a specific procedure comparison and retain counterexamples to broader claims.

The equivalent-basis study confirms that native regularization can materially change forecasts without changing the linear function class. Exact penalty transport restores equality but does not improve the reference estimator; useful superiority over strong simple baselines remains unestablished.

The conditional intervals remain too wide to establish stable superiority or practical equivalence. Their research-information target does not replace prospective confirmation or historical data-availability verification. Independent ownership measurements would be required for crowding identification; ETF trading activity does not supply them. Reliable crash prediction and actionable position limits remain unestablished.

\clearpage
\appendix
\section{Corrected quantity-competition motivation}\label{app:game}
\subsection{Payoffs and Nash equilibrium}
There are $N\geq1$ risk-neutral traders. Each chooses quantity $q_i\geq0$ and faces the same expected pre-impact edge $a>0$. Aggregate quantity is $Q=\sum_{j=1}^Nq_j$ and linear impact has coefficient $\gamma>0$. Expected profit is
\begin{equation}\label{eq:payoff}
 \pi_i(q_i,q_{-i})=q_i(a-\gamma Q).
\end{equation}
This stylized one-period model excludes inventory risk, funding constraints, fees, and heterogeneous information. Its predictions concern this payoff function, not every form of trading competition.

\begin{proposition}[Symmetric equilibrium]
The unique symmetric interior Nash equilibrium of \eqref{eq:payoff} is
\begin{align}
 q^*&=\frac{a}{\gamma(N+1)}, & Q^*&=\frac{Na}{\gamma(N+1)},\\
 e^*&=a-\gamma Q^*=\frac{a}{N+1}, &
 \pi_i^*&=q^*e^*=\frac{a^2}{\gamma(N+1)^2}.
\end{align}
Here $e^*$ is the expected residual edge per unit, not profit per trader or a portfolio Sharpe ratio.
\end{proposition}
\begin{proof}
Holding rivals' quantities fixed, the first-order condition is
\[
 \frac{\partial\pi_i}{\partial q_i}=a-2\gamma q_i-\gamma\sum_{j\neq i}q_j=0.
\]
The second derivative is $-2\gamma<0$. Symmetry gives $a-\gamma(N+1)q^*=0$, and substitution gives the remaining expressions. Positive $a$ ensures an interior symmetric solution.
\end{proof}
Imposing symmetry \emph{before} taking the unilateral derivative changes the optimization problem. The earlier expression $a/(2\gamma N)$ coincides with the Nash quantity only for $N=1$. For $N=2$ and $a=\gamma=1$, the equilibrium is $q^*=1/3$, whereas $1/4$ leaves a positive marginal incentive to increase quantity.

\subsection{Entry and the hyperbolic approximation}
For deterministic effective participation $N(t)=N_0+\lambda t$, the edge is
\begin{equation}
 e(t)=\frac{a}{N_0+1+\lambda t}
     =\frac{K}{1+\widetilde\lambda t},\qquad
 K=\frac{a}{N_0+1},\quad\widetilde\lambda=\frac{\lambda}{N_0+1}.
\end{equation}
Noninteger $N(t)$ is a mean-field approximation to participation. The inverse-linear shape is a prediction for edge under this approximation. Profit per trader instead follows $a^2/[\gamma(N_0+1+\lambda t)^2]$.

\begin{figure}[ht]
\centering\includegraphics[width=.85\linewidth]{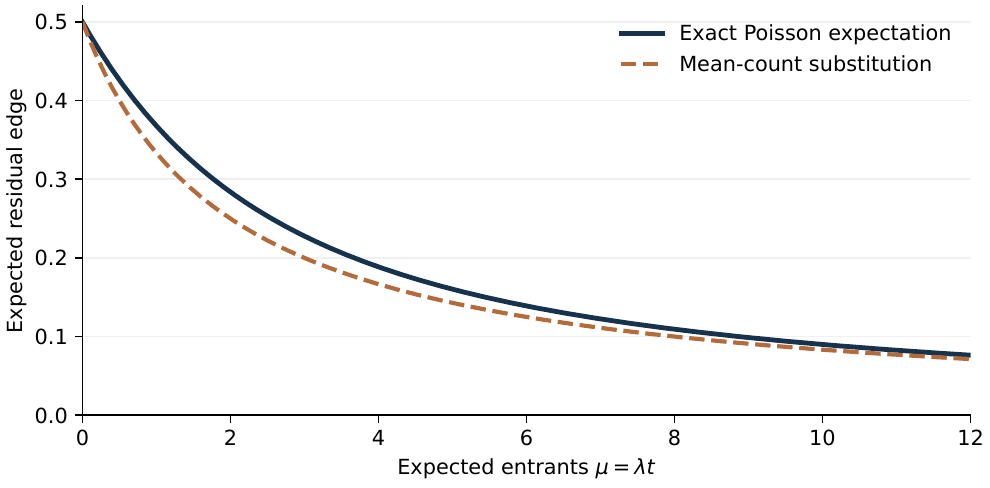}
\caption{Analytical illustration for $a=1$, $N_0=1$. Poisson averaging and substitution of mean participation yield different curves. No factor-return data are used.}\label{fig:entry}
\end{figure}

For stochastic entry, write $N(t)=N_0+M_t$ with $M_t\sim\operatorname{Poisson}(\lambda t)$. The initial count $N_0\geq1$ avoids an empty-market ambiguity.
\begin{proposition}[Exact expected edge under Poisson entry]
With $\mu=\lambda t$,
\begin{equation}\label{eq:poisson}
 \E[e(t)]=a\int_0^1 x^{N_0}\exp\{\mu(x-1)\}\,dx
 \geq\frac{a}{N_0+1+\mu}.
\end{equation}
For $N_0=1$ and $\mu>0$,
\begin{equation}
 \E[e(t)]=a\frac{\mu-1+e^{-\mu}}{\mu^2},\qquad
 \E[e(0)]=a/2.
\end{equation}
\end{proposition}
\begin{proof}
Use $1/(N_0+1+m)=\int_0^1x^{N_0+m}\,dx$ and the Poisson probability-generating function $\E[x^{M_t}]=\exp\{\mu(x-1)\}$. Interchanging expectation and integration is justified by nonnegativity. Jensen's inequality for the convex function $y\mapsto1/y$ gives the lower bound. Direct integration gives the closed form; its limit at zero is $a/2$.
\end{proof}
For $\mu=1$ and $a=1$, the exact mean is $e^{-1}\approx0.36788$, compared with the plug-in value $1/3$. Treating the plug-in curve as exact confuses $\E[1/X]$ with $1/\E[X]$. The analytical curve in Figure~\ref{fig:entry} illustrates this distinction and is not an empirical fit.

\subsection{A finite-entry extension and its limits}
Exogenous Poisson entry is an assumption, not a consequence of the quantity game. To illustrate the difference, consider an alternative two-stage model: identical potential traders decide whether to pay a fixed participation cost $F>0$, then active traders play the quantity game. Let $\pi(N)=a^2/[\gamma(N+1)^2]$ denote gross equilibrium profit with $N\geq1$ active traders.
\begin{proposition}[Weakly stable entry counts]
A positive integer count $N$ supports no profitable exit or entry after quantity re-equilibration if and only if
\begin{equation}
 \frac{a^2}{\gamma(N+1)^2}\geq F\geq
 \frac{a^2}{\gamma(N+2)^2}.
\end{equation}
Equivalently, with $A=a/\sqrt{\gamma F}$, $A-2\leq N\leq A-1$. Zero entry is weakly stable when $F\geq a^2/(4\gamma)$. At equality boundaries two adjacent counts may be stable.
\end{proposition}
\begin{proof}
An incumbent compares $\pi(N)-F$ with zero; an entrant compares $\pi(N+1)-F$ with zero. The two weak inequalities are exactly these incentive constraints. Taking positive square roots gives the stated interval. At zero entry, a deviating entrant becomes a monopolist with gross profit $a^2/(4\gamma)$.
\end{proof}
For $a=\gamma=1$ and $F=0.1$, $N=2$ is stable: $\pi(2)=1/9>0.1$ but $\pi(3)=1/16<0.1$. This model yields finite participation under fixed primitives; it does not derive indefinite hyperbolic decay over calendar time. It assumes participation cost is avoidable at the entry stage and that the quantity subgame is replayed after a deviation. It is not a dynamic exit model with sunk costs.

\subsection{An explicit, restrictive bridge to Sharpe}
Suppose the realized one-period profit of an active trader is $\Pi_i=q_i^*(e^*+\epsilon)$, conditional on fixed $a,N,\gamma$, where $\E[\epsilon]=0$ and $\operatorname{Var}(\epsilon)=\sigma^2>0$. Then
\[
 \frac{\E[\Pi_i]}{\sqrt{\operatorname{Var}(\Pi_i)}}
 =\frac{q_i^* e^*}{q_i^*\sigma}
 =\frac{a}{\sigma(N+1)}.
\]
Thus a population Sharpe ratio inherits the inverse-linear shape if per-unit noise variance is fixed and positions are deterministic conditional on the primitives. This is an added observation model, not a consequence of the expected-profit game. If $\sigma$ varies with participation, even the shape can change; with $\sigma(N)\propto(N+1)^{-1}$ the conditional Sharpe is constant. Random participation adds variation in conditional means, so averaging conditional Sharpe ratios does not give the unconditional Sharpe. Nor does this observation model identify published factor portfolios with equilibrium trader profits.

\subsection{The rolling statistic is a different observation model}
Even the restrictive population-Sharpe bridge does not derive the curve fitted to an overlapping sample statistic. Let $R_j=m_j+\epsilon_j$ within a window of size $W$, with deterministic means $m_j$ and independent mean-zero noises of common variance $\sigma^2$. Then
\begin{equation}
 \E[\bar R_W]=\bar m_W,\qquad
 \E[s_W^2]=\sigma^2+\frac{1}{W-1}\sum_{j=1}^W(m_j-\bar m_W)^2.
\end{equation}
To verify the second equality, expand $\sum_j(R_j-\bar R_W)^2$ into mean and noise terms. Cross terms have expectation zero; the centered-noise sum has expectation $(W-1)\sigma^2$. Thus a changing edge affects both the numerator's historical average and the denominator through within-window mean variation. Moreover, an expected ratio is not the ratio of these expectations. Even under constant Gaussian primitives and $W>2$,
\[
 \E[\widehat S_W]=\sqrt{12}\frac{m}{\sigma}
 \sqrt{\frac{W-1}{2}}\frac{\Gamma((W-2)/2)}{\Gamma((W-1)/2)},
\]
by independence of the sample mean and variance and the inverse moment of a chi-square variable. The multiplicative factor is not one. Accordingly, the fitted rolling-Sharpe hyperbola is a descriptive feature construction; it is not a structurally identified estimate of participation or a derived tail-risk model.

\subsection{What can be identified from a fitted curve?}
A reduced-form fit can estimate a level and a decay rate, but cannot separately identify $a$, $N_0$, and entry intensity without extra information. In fact, a path $e_t=a/(N_t+1)$ generated by changing participation is observationally equivalent in edge to fixed participation $\bar N$ and opportunities $a_t=(\bar N+1)e_t$. Neither return decay nor a good curve fit distinguishes these explanations. The assumptions of constant $a$ and constant $\lambda$ are substantial. Time-varying opportunities, exit, or heterogeneous traders can alter the shape. The impact coefficient $\gamma$ cancels from edge in the stated quantity game; it affects edge indirectly only through extra relationships such as endogenous entry.

The first two moments also do not specify a tail mechanism. At edge $e=0.2$ and unit noise variance, equiprobable noise $\epsilon=\pm1$ gives loss probability $1/2$. Noise taking $-\sqrt{19}$ with probability $0.05$ and $1/\sqrt{19}$ otherwise has the same zero mean and unit variance but loss probability $0.05$. Both have the same population Sharpe. Thus linking participation to tail probabilities requires additional distributional assumptions and independent measurements.

\FloatBarrier
\section{Original evaluation principles and identification requirements}\label{app:protocol}
\subsection{Descriptive decay comparison}
Prespecify constant, linear, exponential, inverse-linear, and flexible trend baselines, with comparable tuning access. Set the time origin before evaluation and report sensitivity to alternative origins as exploratory. Use common signed observations, chronological development folds, and held-out loss differences. Include uncertainty around those differences; in-sample $R^2$ alone does not establish predictive superiority. Factor taxonomy must be defined using external characteristics before fitting, rather than inferred from which factors fit the model.

\subsection{Signal availability and the forecast target}
For the chronological research model, let $\mathcal F_t$ contain observations indexed through month $t$. Real-time use additionally requires the publication-availability restrictions above. Fit the curve using data available before the statistic being predicted to obtain $\widehat m_{f,t|t-1}$. Once month $t$ is observed, define
\begin{equation}
 C_{f,t}=\widehat m_{f,t|t-1}-\widehat S_{f,t}.
\end{equation}
A large positive residual is labeled a \emph{decay-residual signal}. It is not a direct measure of investor holdings or crowding. It can be used to predict month $t+1$, but not described as predicting the already observed return at $t$.

Define a crash outcome $Y_{f,t+1}=\ind\{r_{f,t+1}<q_{f,0.10}^{\rm train}\}$, with the threshold fixed on a documented training window. A recursive threshold is also possible, but all comparisons must use the same outcome definition. For comparability across factors, standardize $C_{f,t}$ using training information only. Preserve non-crash months in the evaluation population.

\subsection{Incremental tail-risk prediction}
Compare a baseline containing lagged factor return, trailing mean return, realized volatility, and a prespecified time trend with the same model augmented by $C_{f,t}$. A logistic specification is
\begin{equation}
 \Prb(Y_{f,t+1}=1\mid\mathcal F_t)
 =\operatorname{logit}^{-1}(b_f+\theta_f C_{f,t}+\eta^\top Z_{f,t}),
\end{equation}
where $Z_{f,t}$ contains only available controls. Small event counts motivate shrinkage or a pooled model with clearly stated pooling assumptions. Evaluate log loss, Brier score, calibration, and incremental performance over the simple controls. A signal derived from trailing returns must beat appropriately flexible versions of those ingredients to demonstrate incremental predictive performance within the tested model classes. This is distinct from adding new information to the full return history.

Also report event counts and denominators for conditional crash rates, risk differences, and risk ratios. Use time-block resampling that preserves contemporaneous dependence across factors, repeating signal fitting when uncertainty in that fit matters. Specify the tested family, including all factors, thresholds, horizons, and signal variants examined. Keep exploratory variants separate from the primary comparison.

\subsection{Independent crowding evidence and trading relevance}
ETF activity can reflect many mechanisms. Correlating cumulative residuals with a trending volume series is vulnerable to common trends. A mechanistic extension should specify the ETF universe, constituent mapping, source, volume or asset measure, transformations, and lag convention; then test incremental associations after trend and market controls. Holdings overlap, flows, and strategy capacity are potential independent measurements, not interchangeable proxies.

For trading, derive positions from information available at the execution time, use identical risk targets across benchmarks, and report turnover, costs, financing, gross and net performance, and uncertainty. A failed strategy cannot prove market efficiency or exclude useful alternative signals. Conversely, a statistical tail association does not by itself justify position limits or stop-loss rules; those decisions require a specified loss function and an evaluation of the resulting policy.

\section{Reassessment of the original reported evidence}\label{app:original}
The original momentum fit reports $R^2=0.65$ for the hyperbolic curve and $0.61$ for the exponential curve, on positive Sharpe observations. These are descriptive values under a selected sample. The reported strategy Sharpe ratios are $0.22$ for crowding timing, $0.39$ for factor momentum, and $0.17$ for equal weighting. They do not establish that crowding is efficiently priced, because signal quality, timing, and policy design remain competing explanations.

\begin{table}[ht]
\centering\small
\begin{tabular}{lrrrr}\toprule
Factor & Crash rate, high signal & Low signal & Reported ratio & $p$\\\midrule
Short-term reversal & 16.9\% & 9.2\% & 1.84 & 0.078\\
Long-term reversal & 19.4\% & 11.8\% & 1.65 & 0.095\\
Momentum & 10.9\% & 28.2\% & 0.38 & 0.006\\\bottomrule
\end{tabular}
\caption{Selected historical values from Table 4 of arXiv:2512.11913v1. The original calls the states crowded/uncrowded. Forecast timing, counts, dependence treatment, and reproduction remain unverified; these are not new results.}
\end{table}
Neither reversal $p$-value is below $0.05$ even before multiple-testing adjustment. The momentum association is more pronounced in the reported table, but its sign and economic interpretation should not be called a verified mechanism. The table gives six factor rows although the study describes eight factors. The omitted rows and complete test family must be restored before correction and interpretation.

The earlier text defines the signal from the current rolling Sharpe and does not make the signal-to-crash lag sufficiently explicit. If the current return contributes to both the signal and the crash event, part of the association can be mechanical. The reconstruction evaluates chronology-correct predictive associations; it cannot attribute the original result to timing without reproducing the original executable analysis. The reported ETF correlation of $-0.63$ also remains descriptive until trend dependence and the measurement protocol are resolved.

\clearpage
\section{Additional reconstruction diagnostics}\label{app:additional}
\begin{table}[!ht]
\centering\small\resizebox{\linewidth}{!}{\begin{tabular}{lrrrrr}\toprule
Factor & Hyperbolic & Constant & Exponential & Linear & Persistence\\\midrule
Mkt-RF & 0.3214 & 0.3214 & 0.3214 & 0.1795 & 0.0294\\
SMB & 0.1545 & 0.2732 & 0.1230 & 0.1160 & 0.0195\\
HML & 0.5951 & 0.7448 & 0.5444 & 0.4677 & 0.0198\\
RMW & 0.2212 & 0.2212 & 0.2212 & 0.3452 & 0.0203\\
CMA & 0.6003 & 0.6822 & 0.5753 & 0.5064 & 0.0206\\
Mom & 0.1650 & 0.3415 & 0.1996 & 0.0900 & 0.0250\\
ST\_Rev & 0.1479 & 0.2923 & 0.1470 & 0.2214 & 0.0226\\
LT\_Rev & 0.8445 & 1.0405 & 0.8369 & 0.6834 & 0.0226\\
\bottomrule\end{tabular}
}
\caption{Mean squared errors for predicting the signed rolling-Sharpe statistic on the common evaluation origins. These are not losses for predicting next-month returns.}
\end{table}
\begin{table}[!ht]
\centering\small\begin{tabular}{rrr}\toprule
Block months & $\Delta$ log loss & 95\% range\\\midrule
6 & +0.00236 & [+0.00038, +0.00450]\\
12 & +0.00236 & [+0.00038, +0.00459]\\
24 & +0.00236 & [+0.00047, +0.00495]\\
36 & +0.00236 & [+0.00061, +0.00456]\\
\bottomrule\end{tabular}

\caption{Block-length sensitivity for the primary macro loss contrast. These descriptive resampling ranges do not establish coverage and do not include full-pipeline refitting or researcher selection.}
\end{table}
\begin{table}[!ht]
\centering\small\begin{tabular}{lrrrr}\toprule
Factor & Same/high & Same/low & Next/high & Next/low\\\midrule
Mkt-RF & 8/44 & 6/76 & 5/43 & 9/77\\
SMB & 10/84 & 5/36 & 8/84 & 7/36\\
HML & 17/98 & 5/22 & 17/97 & 5/23\\
RMW & 14/89 & 6/31 & 12/89 & 8/31\\
CMA & 13/98 & 6/22 & 13/98 & 6/22\\
Mom & 16/98 & 5/22 & 11/98 & 10/22\\
ST\_Rev & 10/50 & 11/70 & 7/50 & 14/70\\
LT\_Rev & 22/84 & 9/36 & 21/84 & 10/36\\
\bottomrule\end{tabular}

\caption{Events/denominator for same-month and next-month state associations on target months 2015--2024. ``High'' means residual above its fixed pre-2015 median; low is the complement. The same-month comparison is diagnostic, not a forecast. Both use the same frozen crash threshold. No inferential claim is made from these secondary counts.}
\end{table}
\clearpage
\subsection{Temporal stability of the fixed primary forecasts}\label{app:temporal}
\begin{table}[!ht]
\centering\small\resizebox{\linewidth}{!}{
\begin{tabular}{lrrrrrrr}\toprule
Period & Events/$N$ & Observed & Controls & Augmented & Prevalence & $\Delta$ & Omit-year $\Delta$\\\midrule
2015 & 15/96 & 0.156 & 0.106 & 0.105 & 0.109 & +0.00099 & +0.00251\\
2016 & 15/96 & 0.156 & 0.110 & 0.112 & 0.109 & -0.00015 & +0.00264\\
2017 & 9/96 & 0.094 & 0.108 & 0.109 & 0.109 & +0.00054 & +0.00256\\
2018 & 13/96 & 0.135 & 0.111 & 0.110 & 0.109 & +0.00267 & +0.00232\\
2019 & 14/96 & 0.146 & 0.111 & 0.112 & 0.109 & +0.00283 & +0.00230\\
2020 & 26/96 & 0.271 & 0.118 & 0.121 & 0.109 & -0.00074 & +0.00270\\
2021 & 12/96 & 0.125 & 0.112 & 0.112 & 0.109 & -0.00012 & +0.00263\\
2022 & 17/96 & 0.177 & 0.119 & 0.118 & 0.109 & +0.00256 & +0.00233\\
2023 & 25/96 & 0.260 & 0.118 & 0.117 & 0.109 & +0.00749 & +0.00179\\
2024 & 17/96 & 0.177 & 0.122 & 0.122 & 0.109 & +0.00749 & +0.00179\\
\midrule
2015-2019 & 66/480 & 0.138 & 0.109 & 0.110 & 0.109 & +0.00138 & ---\\
2020-2024 & 97/480 & 0.202 & 0.118 & 0.118 & 0.109 & +0.00334 & ---\\
All years & 163/960 & 0.170 & 0.113 & 0.114 & 0.109 & +0.00236 & ---\\
\bottomrule\end{tabular}
}
\caption{Post-results descriptive diagnostic. Controls, Augmented, and Prevalence columns report mean predicted probabilities; Observed is the event frequency. $\Delta$ is augmented-minus-control mean log loss in the indicated period. Omit-year $\Delta$ scores the remaining nine years with the original forecasts unchanged. Counts are dependent factor-month events, not independent trials. All years and both fixed halves are retained.}
\end{table}
\FloatBarrier
\subsection{Adaptive fitted-residual null diagnostics}\label{app:nulls}
\begin{table}[!ht]
\centering\small\resizebox{\linewidth}{!}{\begin{tabular}{llrrr}
\toprule
DGP & Cutoff & Same-month RD & Next-month RD & Empty paths \\
\midrule
Gaussian & 50th & 4.3 [-4.8, 13.9] & -1.7 [-14.3, 7.6] & 2/2 \\
Gaussian & 75th & 3.5 [-9.5, 17.0] & -2.5 [-10.9, 6.8] & 7/7 \\
$t_5$ & 50th & 5.0 [-4.1, 15.2] & -0.3 [-10.1, 9.4] & 5/5 \\
$t_5$ & 75th & 6.8 [-7.7, 22.9] & -1.1 [-11.8, 11.4] & 10/10 \\
Centered lognormal & 50th & 3.7 [-6.9, 12.0] & -0.8 [-10.1, 8.2] & 1/1 \\
Centered lognormal & 75th & 4.4 [-7.2, 19.2] & -0.7 [-10.1, 9.2] & 10/10 \\
Stochastic volatility & 50th & 5.9 [-5.3, 19.7] & -1.4 [-12.8, 11.3] & 2/2 \\
Stochastic volatility & 75th & 6.4 [-9.0, 25.6] & -2.6 [-15.4, 12.6] & 4/4 \\
\bottomrule
\end{tabular}
}
\caption{Mean risk differences and 5th--95th between-path percentiles, in percentage points. Each generator has 100 independent paths. Empty paths are same-month/next-month counts with an empty high or low state; these undefined differences are retained in the ledger and omitted only from the explicitly conditional distribution summaries. Bands are not confidence intervals.}
\end{table}
\FloatBarrier
\paragraph{Reproducibility and evidence boundaries.}
The repository contains immutable original PDF, frozen public input archives, a hash-checked parser, configuration, selected penalties and coefficients, generated tables, dated prediction ledgers, and validation scripts. Commands are in the README. The core return analysis and external ETF activity analysis have been rebuilt from public observations. Exact historical replication remains a separate provenance question requiring original files; it is not a prerequisite for executing these new studies. Mechanism identification and certified historical availability still require independent measurements and release-time records. The eight reconstructed factor rows do not resolve omissions from the original reported test family. No original estimates are silently replaced by the reconstructed estimates.

\paragraph{Cross-repository source recovery.}
We inspected the author's \texttt{ai-in-finance} repository, including four available branches and 743 reachable commits. A December 2025 research note explicitly describes this paper's hyperbolic model and factor taxonomy, but the searched history does not supply its eight-factor implementation or archived forecast outputs. The similarly named causal-factor project concerns a different daily regime study. Its available 25-portfolio daily-return cache and return-loading-based co-exposure scores do not constitute the original monthly inputs or independent holdings observations. An IWM-volume download script likewise targets the other study's Granger statistics. The source-recovery audit records commit identifiers, artifact hashes, and remaining requirements; no result from that project is imported as a replication or crowding validation here.

\FloatBarrier
\section{Full regional and matched-history comparisons}\label{app:external}
\begin{table}[!ht]
\centering\small\begin{tabular}{lrrrr}\toprule
 & \multicolumn{2}{c}{Frozen} & \multicolumn{2}{c}{Annual}\\
Model & Log loss & Brier & Log loss & Brier\\\midrule
Prevalence & 0.30787 & 0.08411 & 0.31424 & 0.08494\\
Controls & 0.35215 & 0.08809 & 0.32160 & 0.08507\\
Hyperbolic residual & 0.37651 & 0.08709 & 0.32210 & 0.08501\\
Sharpe & 0.35845 & 0.08819 & 0.32474 & 0.08561\\
Constant residual & 0.35198 & 0.08701 & 0.32465 & 0.08505\\
Linear residual & 0.37253 & 0.08786 & 0.32844 & 0.08547\\
\bottomrule\end{tabular}

\caption{Pooled external scores, January 2025--August 2026 (20 monthly vectors). Secondary-score rankings need not match log-loss rankings.}
\end{table}
\begin{table}[!ht]
\centering\small\begin{tabular}{lllrr}\toprule
Universe & Era & Fit & $\Delta$ log loss & $\Delta$ Brier\\\midrule
Europe & 2015--24 & Frozen & +0.00937 & -0.00211\\
Europe & 2015--24 & Annual & +0.00341 & +0.00004\\
Europe & 2025--26 & Frozen & +0.05476 & -0.00196\\
Europe & 2025--26 & Annual & +0.00299 & +0.00057\\
Japan & 2015--24 & Frozen & +0.00033 & +0.00003\\
Japan & 2015--24 & Annual & +0.00008 & -0.00002\\
Japan & 2025--26 & Frozen & +0.00230 & +0.00003\\
Japan & 2025--26 & Annual & -0.00182 & -0.00024\\
Asia Pacific ex Japan & 2015--24 & Frozen & +0.01141 & +0.00012\\
Asia Pacific ex Japan & 2015--24 & Annual & +0.00349 & +0.00004\\
Asia Pacific ex Japan & 2025--26 & Frozen & +0.01602 & -0.00107\\
Asia Pacific ex Japan & 2025--26 & Annual & +0.00034 & -0.00053\\
US (matched) & 2015--24 & Frozen & -0.04739 & -0.00148\\
US (matched) & 2015--24 & Annual & -0.00608 & -0.00103\\
US (matched) & 2025--26 & Frozen & -0.13071 & -0.00113\\
US (matched) & 2025--26 & Annual & +0.00920 & +0.00192\\
\bottomrule\end{tabular}

\caption{Hyperbolic-minus-controls contrasts for every region, era, and fitting mode. The US benchmark uses the same six-factor common history and split; it is descriptive and excluded from the external confidence family. Negative values favor augmentation.}
\end{table}
\FloatBarrier

\FloatBarrier

\clearpage
\section{Complete equivalent-basis comparisons}\label{app:basis}
\begin{table}[!ht]
\centering\small
\begin{tabular}{lllrrr}
\toprule
Target & Cell & Curve & R--M log loss & R--M Brier & R--S log loss \\
\midrule
Full & A & Hyperbolic & -0.05879 & -0.00510 & -0.00028 \\
Full & A & Exponential & -0.07549 & -0.00505 & -0.00028 \\
Full & A & Linear & -0.10399 & -0.00425 & -0.00020 \\
Full & A & Constant & -0.11571 & -0.00582 & -0.00021 \\
Full & B & Hyperbolic & -0.07174 & -0.00283 & +0.04751 \\
Full & B & Exponential & -0.07029 & -0.00050 & +0.00114 \\
Full & B & Linear & +0.02599 & +0.00083 & -0.00146 \\
Full & B & Constant & -0.00501 & -0.00266 & +0.03541 \\
Full & C & Hyperbolic & -0.06664 & -0.00041 & +0.03012 \\
Full & C & Exponential & -0.07029 & -0.00050 & +0.00114 \\
Full & C & Linear & +0.02599 & +0.00083 & -0.00146 \\
Full & C & Constant & -0.00501 & -0.00266 & +0.03541 \\
Short & A & Hyperbolic & -0.01424 & -0.00398 & +0.00007 \\
Short & A & Exponential & -0.04156 & -0.00439 & +0.00017 \\
Short & A & Linear & -0.04628 & -0.00230 & -0.03540 \\
Short & A & Constant & -0.03073 & -0.00409 & -0.00008 \\
Short & B & Hyperbolic & -0.07941 & -0.00224 & -0.03688 \\
Short & B & Exponential & +0.00986 & +0.00085 & -0.03628 \\
Short & B & Linear & +0.00504 & +0.00017 & -0.03824 \\
Short & B & Constant & -0.03852 & -0.00261 & -0.03566 \\
Short & C & Hyperbolic & -0.03949 & +0.00025 & -0.00734 \\
Short & C & Exponential & +0.00986 & +0.00085 & -0.03628 \\
Short & C & Linear & +0.00504 & +0.00017 & -0.03824 \\
Short & C & Constant & -0.03852 & -0.00261 & -0.03566 \\
\bottomrule
\end{tabular}

\caption{All 24 equal-factor evaluation setups. R denotes the native residual pair $(S,m-S)$, M the native component pair $(S,m)$, and S controls plus Sharpe. Negative differences favor R. Every model selects its numerical penalty on the same validation calendar and refits on development data. Transported residual fits reproduce M, not a generally better predictor. Histories and targets reuse the same evaluation returns; rows are not independent replications.}\label{tab:basis}
\end{table}

\end{document}